\documentclass[journal]{IEEEtran}
\IEEEoverridecommandlockouts

\usepackage[utf8]{inputenc}
\usepackage{setspace}
\usepackage[T1]{fontenc}
\usepackage[american]{babel}
\usepackage{graphicx}
\usepackage{cite}
\usepackage{listings}
\usepackage{subfig}
\usepackage{booktabs}
\usepackage{float}
\usepackage{amsmath,amssymb,exscale}
\newtheorem{lemma}{\bf Lemma}

\usepackage{blindtext, graphicx}
\usepackage{verbatim}
\usepackage{multirow} 
\usepackage{algorithm}
\usepackage{algpseudocode}
\usepackage{fancyvrb}
\usepackage{bera}
\usepackage{amsmath}
\usepackage{mathtools}
\usepackage{lipsum}

\usepackage{times}
\usepackage{scalerel}
\usepackage{tikz}
\usetikzlibrary{svg.path}
\usepackage[bookmarks=false]{hyperref}
\begin{document}
\author{Ruolin Du, Zhiqiang Wei, \IEEEmembership{Member, IEEE,} Zai Yang, \IEEEmembership{Senior Member, IEEE,} Lei Yang, \IEEEmembership{Senior Member, IEEE,} Yong Zeng, \IEEEmembership{Fellow, IEEE,} Derrick Wing Kwan Ng, \IEEEmembership{Fellow, IEEE,} Jinhong Yuan, \IEEEmembership{Fellow, IEEE}
\thanks{
This work is supported in part by the National Natural Science Foundation of China under Grant U24A20214, in part by the Qin Chuang Yuan High-Level Innovation and Entrepreneurship Talent Program under Grant QCYRCXM-2023-094, in part by the National Natural Science Foundation of China under Grants 12526209 and 12371464, and in part by the National Natural Science Foundation of China under Grant 62571116 and the Natural Science Foundation for Distinguished Young Scholars of Jiangsu Province under Grant BK20240070.
An earlier version of this paper was presented at the IEEE ICC 2025 \cite{du2025channel}.

 Ruolin Du and Zai Yang are with the School of Mathematics and Statistics, Xi'an Jiaotong University, Xi'an 710049, China (e-mails: duruolin@stu.xjtu.edu.cn; yangzai@xjtu.edu.cn). 
 
 Zhiqiang Wei is with the School of Mathematics and Statistics, Xi'an Jiaotong University, Xi'an 710049, China, and also with the Peng Cheng Laboratory, Shenzhen, Guangdong 518055, China, and also with the Pazhou Laboratory (Huangpu), Guangzhou, Guangdong 510555, China (e-mail: zhiqiang.wei@xjtu.edu.cn).
 \emph{(Corresponding author: Zhiqiang Wei)}.

 Lei Yang is with the Technology and Engineering Center for Space Utilization, Chinese Academy of Science, Beijing 100094, China (e-mail: yang.lei@csu.ac.cn).

 Yong Zeng is with the National Mobile Communications Research Laboratory, Southeast University, Nanjing 210096, China, and also with the Purple Mountain Laboratories, Nanjing 211111, China (e-mail: yong\_zeng@seu.edu.cn).

 Derrick Wing Kwan Ng and Jinhong Yuan are with the School of Electrical Engineering and Telecommunications, University of New South Wales, Sydney, NSW 2052, Australia (e-mails: w.k.ng@unsw.edu.au; j.yuan@unsw.edu.au).}
 \vspace{-2.5em}
}
\title{Channel Knowledge Map-assisted Dual-domain Tracking and Predictive Beamforming for High-Mobility Wireless Networks}
\IEEEoverridecommandlockouts 
\maketitle

\begin{abstract}
This paper introduces a novel channel knowledge map (CKM)-assisted dual-domain tracking and predictive beamforming scheme for high-mobility wireless networks. 
The central premise is that the CKM integrates both the coordinate and beam domains, thereby enabling tracking in one domain via treating the other domain's input as priors or measurements. 
In the coordinate domain (C-Domain), an extended Kalman filter (EKF) is employed to predict and track the state (i.e., location and velocity) of a moving communication receiver across time slots under both line-of-sight (LoS)-present and LoS-absent conditions, where the CKM provides a prior mapping from multipath channel parameters to potential target locations. 
In the beam domain (B-Domain), the updated location of the receiver is fed back to CKM to offer \textit{a priori} information of angle of arrival (AoA) variations, which are incorporated to establish beam transition models for effective beam tracking, depending on the angular variation situation of each path. 
Then, we analyze the Cramér-Rao Bound (CRB) for AoA estimation for each path in the considered system and propose a jointly predictive beamforming and power allocation design to minimize AoA estimation errors, directly enhancing multipath beam tracking accuracy and indirectly improving target tracking performance.
Simulation results demonstrate that the proposed scheme achieves significant improvements in both target and beam tracking performance compared to the state-of-the-art approaches, particularly in AoA tracking of non-line-of-sight (NLoS) paths, highlighting the potential gain of CKM in facilitating both target and beam tracking in high-mobility communications.
\end{abstract}

\begin{IEEEkeywords}
Tracking, dual-domain, channel knowledge map, extended Kalman filter, high-mobility, predictive beamforming.
\end{IEEEkeywords}

\vspace{-1em}
\section{Introduction}

\subsection{Background}
The upcoming sixth-generation (6G) wireless networks are expected to achieve unprecedented data rates, increased connection density, and superior reliability \cite{zengEnvironmentAware6GCommunications2021}. 
A critical aspect of 6G evolution lies in supporting high-mobility scenarios, which are vital for emerging applications involving unmanned aerial vehicles (UAVs), ground vehicles, and high-speed trains \cite{Wei2023UAV,zhaoSensingAssistedPredictiveBeamforming}. 
In practice, these high-mobility scenarios introduce formidable challenges for accurate channel estimation or prediction due to the rapid time-varying channel conditions, further complicated by the expansion of the channel matrix across both frequency and spatial domains. 
The task of estimating these dynamic and large-scale channel matrices is computationally-intensive and requires a large amount of channel training overhead. 
Therefore, various advanced techniques such as beam tracking and predictive beamforming (BF) are essential to maintain high-quality links in high-mobility wireless networks, enabling real-time system adaptation to dynamic channel variations \cite{liISACEnabledV2INetworks2023,zhangFastBeamTracking2019,zhaoSensingAssistedPredictiveBeamforming,cui2024SeeingNotAlwaysBelievingISACAssistedPredictiveBeamTrackingMultipathChannels}.
Moreover, integrated sensing and communication (ISAC) has been recognized as one of the six key application scenarios for 6G \cite{ITU}.
In fact, leveraging communication signals for target tracking in high-mobility wireless networks not only meets emerging demands for localization services but also improves communication performance, particularly when the moving target is also the communication terminal.

One of the key challenges in high-mobility wireless networks is tracking the angles of arrival (AoA) of moving communication terminals, such as vehicles in vehicle-to-everything (V2X) networks.
%
In particular, multipath AoA information is essential for both channel reconstruction and wireless positioning. 
However, AoA tracking in multipath environments is much more complicated than that in line-of-sight (LoS) propagation environments, particularly for non-line-of-sight (NLoS) paths. 
Specifically, NLoS paths typically experience higher propagation loss than that of the LoS path, leading to reduced receiving signal strength and AoA estimation accuracy. 
Moreover, the rapid changes in the AoA, driven by vehicle maneuvers and the occurrence of random obstacles, pose significant challenges for tracking multipath AoA. 
In addition, the fast AoA variation leads to a severe model mismatch between the assumed beam transition model and the ground truth. 
Even worse, the random path birth and death, or burst variations of multipath AoA significantly reduce the temporal correlation of beam transition and thus might cause the traditional beam tracking algorithm to fail \cite{zengCKMAssistedIdentificationPredictive2023a}. 
As a result, incorporating extra environment information is essential to ensure reliable and accurate beam tracking in multipath environments, yet it has not been fully exploited.

Target tracking has long been a cornerstone of radar technology research \cite{zhangFastBeamTracking2019,zhangTrackingAnglesDeparture2016a}, and has recently regained significant interest in the wireless communication sector due to the emerging of the ISAC paradigm.
Traditionally, radar-based target tracking usually assumes LoS propagation and ignores multipath effects, as radar typically operates in open-air spaces with extensive sensing ranges.
However, ISAC operates on wireless infrastructures in complicated propagation environments, such as urban, where the sensing range is usually limited within a cell, say 500 meters \cite{500m}.
Therefore, multipath effects and the random absence of the LoS path are inevitable for target tracking in the context of ISAC.
Unfortunately, when the LoS path is obscured, the NLoS paths, caused by scatterers in the propagation environment, are not directly related to the moving target, thus complicating their explicit exploitation for target tracking.
In this case, acquiring comprehensive environmental information becomes essential for effective beam tracking and target tracking in NLoS propagation environments.

Channel knowledge map (CKM) is a novel technique connecting environmental information and channel state information (CSI) for a specific area \cite{zengEnvironmentAware6GCommunications2021, zengTutorialEnvironmentAwareCommunications2023,xu2024how}. 
Beyond facilitating an understudy of channel temporal/frequency correlations, the CKM provides more channel \textit{a priori} information by mapping geographic positions to CSI, which enables more accurate and robust beam tracking performance \cite{wu2024environment}, especially in NLoS propagation environments. 
However, the challenge remains that location information for a moving terminal is usually unavailable at the transmitter, and the method to effectively integrate channel temporal correlations and the \textit{a priori} channel information provided by the CKM remains unclear.
Furthermore, the CKM can also be inversely exploited to improve target localization and tracking performance by mapping the estimated CSI to geographic \textit{a priori} information.
In particular, for the case with the absence of LoS path, CKM can provide geographic \textit{a priori} information via exploiting the estimated NLoS path parameters and thus might improve sensing performance in NLoS propagation environment.
However, obtaining accurate multipath information in high-mobility wireless networks is itself a challenging task.
Therefore, this paper aims to propose a dual-domain tracking scheme, which simultaneously performs target tracking in the coordinate domain (C-Domain) and beam tracking in the beam domain (B-Domain), while bridging these two domains through the CKM.

In high-mobility wireless networks, predictive BF is critical to maintaining consistent link quality, as the procedures for CSI acquisition and BF design require substantial training overhead.
In particular, based on predicted CSI or multipath AoA information, the transmitter designs BF strategies preemptively that optimize system communication and/or sensing performance for upcoming time slots.
However, most existing works on predictive BF algorithms assume the presence of a LoS path \cite{shahamFastChannelEstimation2019}, a condition that often does not hold in complex environments, such as urban vehicular networks.
In these settings, LoS path might be obstructed by tall buildings or other vehicles and NLoS paths are typicality inevitable.
Moreover, as analyzed in \cite{cui2024SeeingNotAlwaysBelievingISACAssistedPredictiveBeamTrackingMultipathChannels}, BF directly towards the AoA of the LoS path is not always the optimal strategy in terms of achievable rate, with a performance gap of nearly one-third compared to optimal BF in a multipath channel.
Indeed, effective predictive BF design in NLoS propagation environments enables channel utilization and provides a stronger received signal/echo power, which in turn improves the beam tracking performance.
Building on this insight, this paper further designs a predictive BF strategy that provides high-quality measurements for the proposed CKM-assisted dual-domain tracking scheme.

\subsection{Existing Works}

\label{existingworks}
Beam tracking has been extensively studied in literature for high-mobility wireless networks, though most existing studies have relied on LoS-dominant channel model.
To mitigate the substantial overhead, which can reach up to $43.24\%$ in fifth-generation (5G) New Radio (NR) networks \cite{liISACEnabledV2INetworks2023}, the echo of the downlink signal reflected from the moving terminal is directly processed to estimate its state parameters (e.g., range, velocity, AoA, etc.). 
For instance, in \cite{zhangFastBeamTracking2019}, the authors proposed a beam tracking algorithm leveraging the temporal correlation of AoA, modeling the AoA variation within a single transmission frame as a discrete Markov process. 
Furthermore, in \cite{zhangTWC2019codebook}, an ``on-grid'' AoA transition model was adopted for simplicity, assuming that the AoA of each path shifts among some predetermined angular grid points. 
Additionally, in \cite{seoTrainingBeamSequence2016a}, the beam tracking problem was formulated as a Markov decision process and the corresponding training beam sequence was optimized by exploiting a reinforcement learning (RL) framework, albeit at a high computational cost.
However, the adopted narrow training beams \cite{seoTrainingBeamSequence2016a} are inadequate for tracking rapid beam variations in high-mobility scenarios. 
To overcome this, multiple radio frequency (RF) chains were advocated to generate multiple probing beams for improving tracking performance.
On the other hand, for target tracking, in \cite{liuRadarAssistedPredictiveBeamforming2020a}, an extended Kalman filter (EKF) framework was proposed to track the location and angular variation of a vehicle, based on echo signal measurements and the vehicle's state evolution model. 
In \cite{risticTutorialBernoulliFilters2013a}, the authors proposed a Bernoulli Gaussian sum filter for tracking dynamic targets that randomly appear and disappear, especially in multi-sensor scenarios.
Additionally, the unscented Kalman filter (UKF) \cite{UKF} and the particle filter \cite{ParticleFilter} have also been proposed for tracking moving targets in high-mobility wireless networks.

In practice, the LoS path is not always available and NLoS paths are more frequently encountered in wireless channels, which imposes challenges for both beam and target tracking, as previously discussed.
Although few works have studied the beam and target tracking problems in NLoS propagation environments \cite{zengCKMAssistedIdentificationPredictive2023a,zhaoSensingAssistedPredictiveBeamforming,2010NLoStargetTOAtracking,2020NLoSTrajectoryandRangingOffset,wangDeepLearningEnabled2021}, research remains limited.
In \cite{zengCKMAssistedIdentificationPredictive2023a,zhaoSensingAssistedPredictiveBeamforming}, CKM was adopted to assist LoS path detection, which improves the robustness of target tracking via disabling state update when the LoS path is absent.
Also, in \cite{2010NLoStargetTOAtracking}, the authors leveraged the probability density function of the state of a moving target to mitigate NLoS propagation effects, employing an EKF to track the target from identified LoS time-of-arrival (TOA) measurements.
%
Furthermore, in \cite{2020NLoSTrajectoryandRangingOffset}, the authors proposed a target tracking scheme for NLoS environments by jointly estimating the target trajectory and sensor-target distance via multi-sensor information fusion.
Moreover, in \cite{wangDeepLearningEnabled2021}, deep learning (DL) was applied to forecast AoAs in NLoS channels, which requires a large amount of data and incurs an expensive computation load due to the need for neural network training.
The primary disadvantages of existing beam tracking methods in NLoS propagation environments include poor performance due to high propagation loss and the impracticality of the commonly-used NLoS model, which neglects the dynamic appearance and disappearance of NLoS paths caused by random obstacles.

Predictive BF has emerged as a key strategy to mitigate frequent link blockages and reduce significant beam training overhead in high-mobility communication systems.
In practice, moving communication terminals in such wireless networks often traverse fixed routes, such as high-speed rails or expressways, providing predictable position and velocity, thus making predictive BF feasible.
For example, in \cite{chenMassiveMIMObeamforming2017}, the authors introduced a low-complexity predictive BF scheme featuring transmit diversity, while the BF vector associated with the predictive location of a moving communication receiver is precalculated and pre-stored at the base station.
In contrast, \cite{tianVisionAidedBeamTracking2021} introduced a vision-aided approach leveraging camera images and DL to predict the optimal beam indices in the next few time slots.
Moreover, in \cite{zhangDeeplearningpredictive2023TVT}, the authors trained DL models based on historical CSI to forecast future beam patterns. 
However, both approaches in \cite{tianVisionAidedBeamTracking2021} and \cite{zhangDeeplearningpredictive2023TVT} require a large amount of data and a high computational demand for training DL models.

Generally, the construction of CKM involves two main stages: data acquisition and map generation \cite{zengTutorialEnvironmentAwareCommunications2023}.
In the data acquisition stage, location-specific channel data can be obtained either from simulation-based methods (e.g., ray tracing over physical environment maps \cite{Lim2020MapBased}) or from measurement-based approaches, such as offline drive tests and online data collection enabled by the minimization of drive tests (MDT) framework standardized in 3GPP \cite{Johansson2012MDT}.
In the map generation stage, model-free interpolation and model-assisted learning constitute the two main categories of methods for reconstructing \cite{Kim2011KrigedKF,ChilesDelfiner2009Geo,CoverHart1967NN,liChannelKnowledgeMap2022,Chouvardas2016ICASSP,Hamilton2013Tomography}. 
Model-free methods, such as Kriging \cite{Kim2011KrigedKF}, inverse-distance-weighted (IDW) interpolation \cite{ChilesDelfiner2009Geo,CoverHart1967NN,liChannelKnowledgeMap2022}, and low-rank matrix completion \cite{Chouvardas2016ICASSP} were widely adopted in recent literature as basic techniques for CKM generation.
Model-assisted approaches construct the CKM by integrating data-driven learning with geometric models of the propagation environment \cite{Hamilton2013Tomography}, leveraging ray-tracing-like techniques to infer environmental characteristics with enhanced accuracy and scalability.
Compared with more complex Kriging or data-driven learning methods, the IDW method offers an excellent trade-off between the interpolation accuracy and computational scalability, which is essential for large-scale high-mobility simulations.

Wireless CKM-assisted sensing expands the design possibilities by providing an interface between beam and coordinate domains. 
In \cite{longEnvironmentAwareWirelessLocalization}, the authors proposed a LoS-map-assisted anchor selection scheme to minimize the Bayesian Cramér-Rao Lower Bound (BCRLB) for ranging.
In addition, the authors in \cite{xuCKMcluttersuppression2024} proposed a clutter angle map (CLAM) to eliminate the clutter-related components from the sensing signal before target detection, thereby enhancing target sensing performance.
To the authors' best knowledge, CKM has not been fully exploited to assist effective beam tracking in multipath channels but has been exploited in the opposite way to improve target tracking performance.
Indeed, CKM connects the C-Domain and B-Domain, potentially improving tracking performance in both domains by treating the information from one domain as measurements for the other.
This integration motivates the proposed dual-domain tracking scheme in this paper.

\subsection{Our Contributions}

To tackle the challenges posed by multipath propagation and high-mobility networks, this paper proposes a CKM-assisted dual-domain tracking framework. 
The proposed scheme simultaneously performs target tracking in the C-Domain and beam tracking in the B-Domain by modeling the state evolution in one domain and utilizing measurements/\textit{a priori} information from the other domain. 
To further improve beam tracking performance, this paper proposes a predictive BF design to minimize the maximum AoA estimation errors across multiple paths. 
Our main contributions are summarized as follows:

\begin{itemize}
    \item We propose a CKM-assisted dual-domain tracking framework that effectively integrates information from both the coordinate and beam domains. 
    In the C-Domain, the location and velocity of the target are the states to be updated, with the estimated channel parameters from the B-Domain acting as measurements, directly for LoS-present cases and indirectly via CKM for LoS-absent cases. 
    In the B-Domain, the received echo serves as the measurement and the multipath AoAs are the state to be updated, via combining the \textit{a priori} information provided by the C-Domain through CKM.
    This integration can enhance B-Domain tracking accuracy by exploiting CKM to reference the tracked state of the moving target and improve target tracking accuracy by inversely leveraging CKM with the tracked CSI.
     
    \item In the C-Domain, we propose an EKF-based target tracking algorithm to predict and track the state of a moving user terminal, with state evolution and measurement models defined for both LoS-present and LoS-absent conditions. 
    Even when the LoS path is obscured, the proposed approach can maintain continuous target tracking by exploiting NLoS path information through the CKM. 
    In the B-Domain, depending on the condition of the beam transition, we propose an adaptive \textit{a priori} information fusion strategy for creating the beam transfer probability matrices (TPMs) to assist B-Domain tracking, which effectively combines the channel temporal correlation and the channel \textit{a priori} information provided by CKM.
    The proposed approach can track the fast and even burst changes of multipath AoA due to the dynamic environment by exploiting CKM.
    
    \item Building upon the predicted AoA, we analyze the Cramér-Rao Bound (CRB) of AoA estimation for each path to accurately characterize the AoA estimation error. 
    Then, a predictive BF and power allocation design is formulated as an optimization problem to minimize the maximum AoA estimation error among multiple paths for the next time slot. 
    We propose a suboptimal approach that first selects beams based on the predicted AoAs and then designs the power allocation among multiple beams via solving a convex optimization problem.
    
    \item We conduct extensive simulations to evaluate the superior performance of the proposed dual-domain tracking and predictive BF scheme. 
    Our simulation results demonstrate that the proposed scheme achieves both accurate and continuous target and beam tracking even in complex high-mobility wireless networks, while traditional tracking schemes often fail. 
    Besides, the proposed predictive BF and power allocation design further improves the B-Domain tracking accuracy, particularly for NLoS paths.
\end{itemize}

The remainder of this article is organized as follows: Section II introduces the system model. 
Section III presents the proposed CKM-assisted dual-domain tracking framework. 
Sections IV and V describe the proposed C-Domain and B-Domain tracking algorithms, respectively. 
Section VI presents a predictive BF design. 
Section VII presents the numerical results and Section VIII concludes the paper.

Unless otherwise specified, matrices are denoted by uppercase bold letters and vectors are represented by lowercase bold letters; 
$\operatorname{vec}(\cdot)$ denotes the vectorization operations; 
$(\cdot)^T$, $(\cdot)^H$, and $(\cdot)^*$ stand for transpose, Hermitian transpose, and the complex conjugate of a matrix; 
$\boldsymbol{I}_N$ denotes the $N$-dimensional identity matrix;
$\boldsymbol{1}_N$ denotes the $N$-dimensional all-ones vector;
$\delta(\cdot)$ denotes the Dirac delta function; $\mathbb{C}$ denotes a complex space; 
$\mathbb{E}$ denotes the expectation operator; 
$\mathrm{Var}(\cdot)$ and $\mathrm{Cov}(\cdot)$ denote the operator to determine the variance and covariance, respectively; 
$\operatorname{diag}(\cdot)$ denotes the diagonal operator;
$\hat{x}$ denotes the estimation of variable $x$; 
$\otimes$ denotes the Kronecker product; 
$\propto$ indicates that the left hand side is proportional to the right hand side; 
$\boldsymbol{x}\sim \mathcal{CN}(\boldsymbol{\mu},\boldsymbol{\Sigma})$ denotes a circularly symmetric complex Gaussian vector with mean $\boldsymbol{\mu}$ and covariance matrix $\boldsymbol{\Sigma}$; 
$\boldsymbol{x}\sim \mathcal{N}(\boldsymbol{\mu},\boldsymbol{\Sigma})$ denotes a real-valued Gaussian vector with mean $\boldsymbol{\mu}$ and covariance matrix $\boldsymbol{\Sigma}$; 
$\boldsymbol{A}_{k,:}$, $\boldsymbol{A}_{:,l}$ and $\boldsymbol{A}_{k,l}$ represent the $k$-th row, $l$-th column and $(k,l)$-th element of $\boldsymbol{A}$; 
$l_1$ norm, $l_2$ norm, and the Frobenius norm are denoted by $\|\cdot\|_1$, $\|\cdot\|_2$, and $\|\cdot\|_F$, respectively.

\section{System Model}

\label{system model}
\begin{figure}[!t]
\centering
\includegraphics[width=3 in]{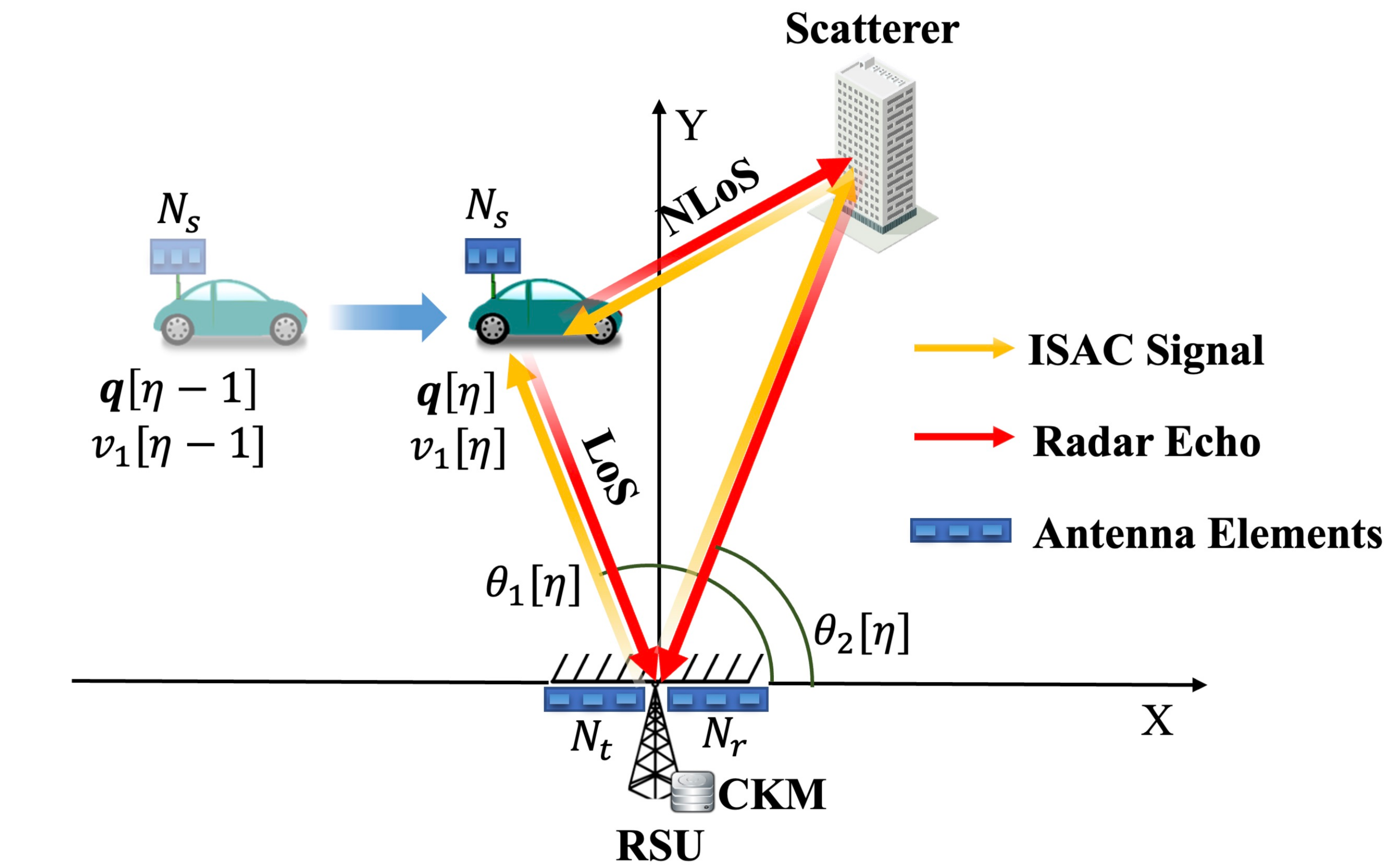}
\caption{Beam tracking system model for a high-mobility vehicle network in multipath environments.}
\label{figsystemmodel}
\end{figure}

Consider a downlink high-mobility wireless vehicular network\footnote{Note that the considered vehicular network is a typical high-mobility network and the proposed dual-domain tracking framework can be extended to other high-mobility wireless networks, such as UAV networks.} as depicted in Fig. \ref{figsystemmodel}. 
A roadside unit (RSU) serves as both a communication transmitter and a sensing receiver, providing services to a moving vehicle. 
The full-duplex operation RSU is equipped with $N_t$ transmitting antennas and $N_r$ receiving antennas, both constituting uniform linear arrays (ULAs) with half wavelength antenna spacing. 
The vehicle is equipped with a ULA of $N_s$ antenna elements for receiving signals from the RSU. 
For ease of exposition, we assume that the vehicle travels along the road parallel to the antenna array of RSU, moving initially towards and subsequently away from the RSU\footnote{To facilitate the presentation of the proposed dual-domain tracking framework, we consider a simple tracking scenario in this paper, as adopted in the literature \cite{liuRadarAssistedPredictiveBeamforming2020a,zengCKMAssistedIdentificationPredictive2023a,liISACEnabledV2INetworks2023,cui2024SeeingNotAlwaysBelievingISACAssistedPredictiveBeamTrackingMultipathChannels}. However, it is worth noting that the proposed framework can be extended to more complicated tracking scenarios. 
For non-parallel linear movement, the velocity of the vehicle can be represented as a two-dimensional vector with components along the x and y axes, requiring only modifications to the state evolution model and measurement models in the C-Domain.  
For curved trajectories, the short slot duration allows adopting a constant velocity model between consecutive slots, where the proposed tracking framework is still applicable.}.
In the considered system model, the position of the vehicle is represented by $\boldsymbol{q}=[q_x,q_y]$ and its velocity is denoted as $v_1$. 
We assume that the velocity is a scalar, with its sign indicating movement direction along the x-axis, defined as the horizontal axis in Fig. \ref{figsystemmodel}.

Assume $T_{\max}$ being the maximum time duration of interest. The time period $T_{\max}$ is divided into several time slots of length $\Delta T$, with each slot further divided into $L$ symbol periods. 
Within each slot, we assume that the channel parameters remain unchanged \cite{liuRadarAssistedPredictiveBeamforming2020a,liISACEnabledV2INetworks2023,yuanIntegratedSensingCommunicationAssisted2021a}. 
The transmitted $N_s$ data streams from the RSU in a slot are given by $\boldsymbol{S}=[\boldsymbol{s}^T_1;\cdots;\boldsymbol{s}^T_{N_s}]=[
 \boldsymbol{x}_1 ,\cdots, \boldsymbol{x}_L
] \in \mathbb{C}^{N_s\times L} $, where $\boldsymbol{s}_{n_s}\in \mathbb{C}^{L \times 1}$, ${n_s}\in \{1,\ldots,N_s\}$, denotes the $n_s$-th downlink data streams and $\boldsymbol{x}_l\in \mathbb{C}^{N_s \times 1}$, denotes the transmitted signal at the $l$-th snapshot, $\forall l\in \{1,\ldots,L\}$.
We assume $\boldsymbol{S} _{i,j}\sim \mathcal{CN}(0,1)$, $\forall i,j$, are i.i.d., leading to $\mathbb{E}[ \boldsymbol{x}_{i}\boldsymbol{x}_{j}^H] = \delta(i-j)\boldsymbol{I}_{N_s}$, $\forall i,j$.
Owing to the independence among different data streams, for sufficiently large $L$ we have $\boldsymbol{s}^H_{n_s}\boldsymbol{s}_{n_s'} \approx \delta( n_s - n_s') L$.
The $N_s$ downlink data streams transmitted at the $[\eta]$-th slot are denoted as $\boldsymbol{S}[\eta]=[ \boldsymbol{x}_1[\eta],\cdots,\boldsymbol{x}_L[\eta] ]\in \mathbb{C}^{N_s\times L}$. 
The precoded signal in the $[\eta]$-th slot at the RSU is $\tilde{\boldsymbol{S}}[\eta]=\boldsymbol{F}[\eta|\eta-1]\boldsymbol{S}[\eta]\in \mathbb{C}^{N_t\times L}$, where $\boldsymbol{F}[\eta|\eta-1]\in \mathbb{C}^{N_t \times N_s}$ is the predictive BF matrix.
In high-mobility scenario, obtaining perfect CSI at the transmitter is challenging. 
Thus, the predictive BF matrix $\boldsymbol{F}[\eta|\eta-1]$ is designed based on the predicted CSI in the proposed dual-domain tracking framework.

Assuming there are a finite number of scatterers between the transmitter and receiver, the received echoes at the RSU consist of the signals reflected by both scatterers and the moving vehicle. 
The sensing channel in the $\eta$-th slot is given by 
\begin{equation}
\small
\label{ChannelModel}
    \boldsymbol{H}[\eta ](t, \tau)=\sum_{i=1}^P \beta_i[\eta ] \boldsymbol{b}(\theta_i[\eta ]) \boldsymbol{a}^{H}(\theta_i[\eta ]) \delta(\tau-\tau_i[\eta ]) e^{j 2 \pi \mu_i[\eta ] t},
\end{equation}
where $i\in \{1,\dots, P\}$ is the path index\footnote{For the LoS-present case, \textbf{path 1} is a LoS path, while \textbf{path 2} is an NLoS path.
For LoS-absent case, \textbf{path 1} becomes the strongest NLoS path.}, $\beta_i[\eta ]\in \mathbb{C}$, $\tau_i[\eta ]\in \mathbb{R}$ and $\mu_i[\eta ]\in \mathbb{R}$ denote the channel path gain, the round-trip delay, and the round-trip Doppler frequency shift of the $i$-th path in the $[\eta]$-th slot, respectively. 
$\boldsymbol{a}(\theta[\eta ])$ and $\boldsymbol{b}(\theta[\eta ])$ denote the steering vectors of the RSU transmitting and receiving ULAs, respectively, and they are defined by $\boldsymbol{a}(\theta[\eta ])= [1, e^{j\pi \cos (\theta[\eta ]) }, \cdots, e^{j\pi(N_t-1) \cos (\theta[\eta ])}]^T \in \mathbb{C}^{N_t\times 1}$, and $\boldsymbol{b}(\theta[\eta ])= [1, e^{j\pi \cos (\theta[\eta ])}, \cdots, e^{j\pi(N_r-1) \cos (\theta[\eta ])}]^T\in \mathbb{C}^{N_r\times 1}$.
The channel path gains $\beta_i[\eta ]$ captures both effects of the radar cross-section (RCS) of the vehicle/scatterers and the propagation loss. 
Denote $\theta_i[\eta ]$ as the angle of the $i$-th path relative to the RSU and we assume that the AoA and angle of departure (AoD) of the $i$-th path at the RSU in the $[\eta]$-th slot are identical, as commonly adopted in \cite{yuanIntegratedSensingCommunicationAssisted2021a,liuRadarAssistedPredictiveBeamforming2020a}.

In the remaining part of this paper, the slot index is omitted without causing ambiguity. 
Accordingly, the echo signal, $\boldsymbol{R}$, received over the $L$ symbol periods at the RSU is given by: 
\begin{equation}
\small
\label{R}
 \boldsymbol{R}= 
\sum_{i=1}^P \beta_i \boldsymbol{b}(\theta_i) \boldsymbol{a}^H(\theta_i) \boldsymbol{F} {\boldsymbol{S}_{l_i}} 
{\boldsymbol{\Lambda}_{k_i}}
+\boldsymbol{Z} \in \mathbb{C}^{N_r \times [L+\max(l_i)]},
\end{equation}
where ${\boldsymbol{S}_{l_i}}\triangleq
[\boldsymbol{0}_{N_s\times l_i} ,\boldsymbol{S} , \boldsymbol{0}_{N_s\times [\max (l_i)-l_i]}] \in \mathbb{C}^{N_s \times (L+\max(l_i))}$ denotes the delayed signal associated with the $i$-th path, $l_i=\frac{\tau_i}{T_p}$ represents normalized delay index within a slot and $T_p=\frac{\Delta T}{L}$ denotes the time domain sample interval.
We assume that the delays are integer multiples of the sampling interval and there is no inter-symbol interference among consecutive time slots by inserting a proper temporal guard interval at the beginning of each time slot \cite{liuRadarAssistedPredictiveBeamforming2020a}.
${\boldsymbol{\Lambda}_{k_i}}$ is the Doppler frequency shift matrix and it is defined by ${\boldsymbol{\Lambda}_{k_i}} \triangleq \operatorname{diag}(\boldsymbol{0}_{l_i}, \boldsymbol{\Lambda}^{\text{shift}}_{k_i}, \boldsymbol{0}_{\max (l_i)-l_i}) \in \mathbb{C}^{(L+\max(l_i)) \times (L+\max(l_i))}$, where $k_i= \mu_i T_p$ represents the normalized Doppler shift and $\boldsymbol{\Lambda}^{\text{shift}}_{k_i} = \operatorname{diag}(1, \dots, e^{j 2\pi k_i (L-1)}) \in \mathbb{C}^{L\times L}$. 
The noise signal is denoted as $\boldsymbol{Z}$ where each entry follows i.i.d. Gaussian with zero mean and variance $\sigma_z^2$, i.e., $\boldsymbol{Z}_{n_r,:}\sim \mathcal{CN}(\boldsymbol{0},\sigma_{z}^2\boldsymbol{I}_{L+\max(l_i)}),\forall n_r=1,\ldots,N_r$, and $\sigma_{z}^2$ is the noise power.

CKM aims to provide location-specific channel parameter tuples for an area of interest.
Specifically, CKM is essentially a mapping function from the user terminal location $\boldsymbol{q}$ to the location-specific channel knowledge such as path gain, angles, delay, and Doppler shift\footnote{
In the CKM literature \cite{zengTutorialEnvironmentAwareCommunications2023}, the channel Doppler shift can be obtained from a CKM based on the location and velocity information.
In practice, the Doppler information can be sampled in a dynamic environment or inferred based on the sampled Doppler shift in a static scattering environment.
This paper focused on tracking algorithm design and thus the CKM construction is out of scope.
Note that the proposed dual-domain tracking scheme can be straightforwardly extended to the case without Doppler information in the CKM.}\cite{zengTutorialEnvironmentAwareCommunications2023}, i.e.,
\begin{equation}
\label{CKM}
\mathcal{M}( \boldsymbol{q},v_1 )=\left \{(\alpha_i,{\theta}_i,\tau_i,\mu_i)|i=1,\cdots,P \right \}.
\end{equation}
where $\alpha_i\in \mathbb{R}$ captures the propagation power loss of the $i$-th path relative to the RSU. 
The channel path gain is given by $\beta_i = \varepsilon \alpha_i^2$, and $\varepsilon$ is the time-invariant reflection coefficient of the considered user terminal \cite{zhangFastBeamTracking2019,liuRadarAssistedPredictiveBeamforming2020a,zhaoSensingAssistedPredictiveBeamforming}.
It is worth noting that, in practice, the CKM is constructed from sampled data using interpolation techniques (as discussed in Section I-B). 
Consequently, the channel parameters stored in the constructed CKM may deviate slightly from the ground-truth values, thereby introducing CKM construction errors.
To facilitate the presentation in sequel, we assume that the number of paths stored in CKM is the same as the number of paths $P$ in \eqref{ChannelModel}. 
In practice, the number of paths during the time duration of interest might be time-varying and be inconsistent with the number of paths stored in the CKM, requiring further investigation, which is left for future work.

\section{CKM-assisted Dual-domain Tracking Framework}

\begin{figure*}[!t]
\centering
\includegraphics[width=6 in]{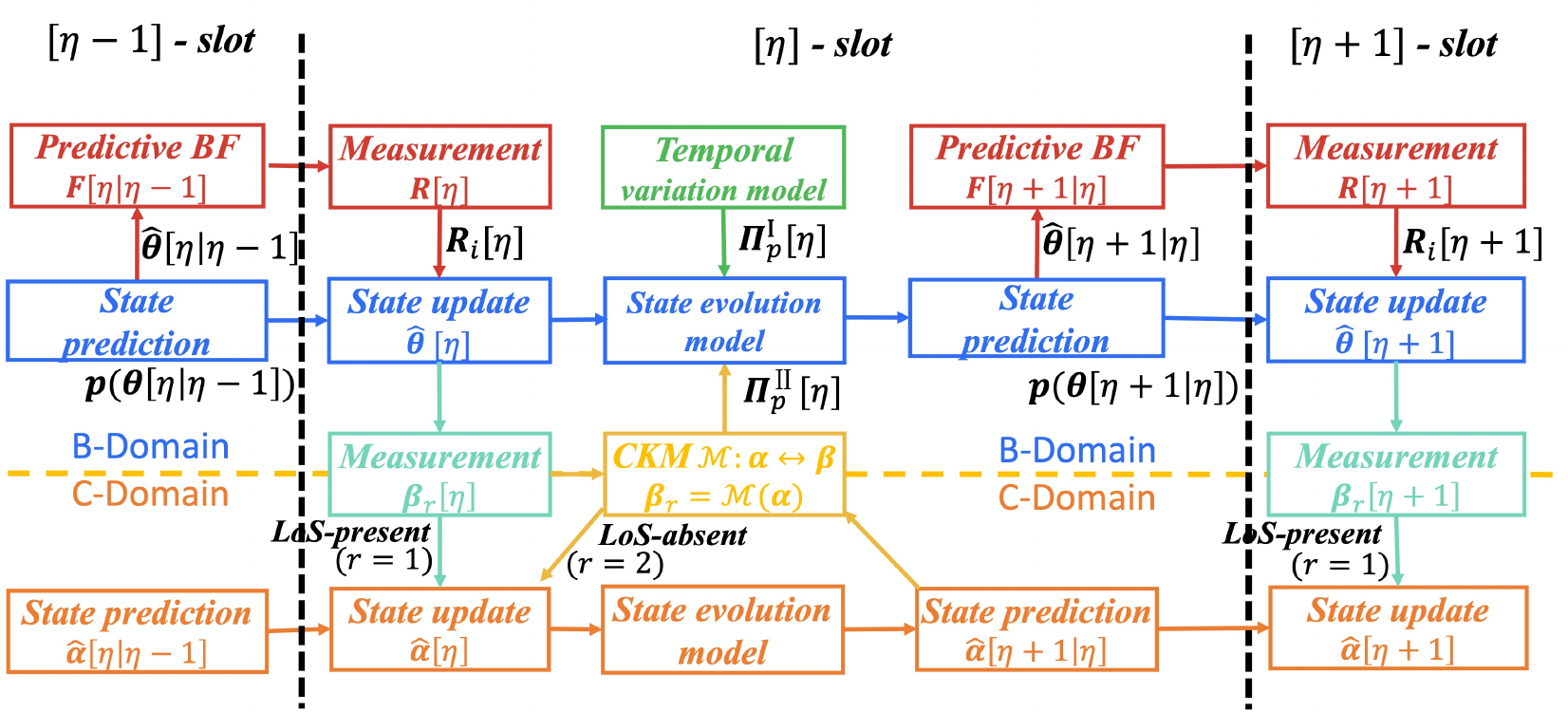}
\caption{CKM-assisted dual-domain tracking framework.}
\label{fig3}
\end{figure*}

Before delving into the details, in this section, we present a high-level overview to the proposed CKM-assisted dual-domain tracking framework, as illustrated in Fig. \ref{fig3}. 
The proposed framework is well-designed with the traditional Kalman filtering framework \cite{liuRadarAssistedPredictiveBeamforming2020a,Wei2023UAV,Wu2023IOT}, which consists of state prediction and update steps across consecutive time slots. 
Distinctively, unlike existing ones, it is defined in both the beam and coordinate domains, interconnected through the CKM.
The proposed dual-domain framework aims to establish a CKM-assisted tracking framework that jointly improves both C-Domain and B-Domain tracking accuracy under both LoS-present and LoS-absent conditions.
Specifically, the C-Domain tracking aims to maximize the posterior probability of the vehicle state by jointly exploiting the C-Domain prior from the previous slot and the current B-Domain measurements, as will be detailed in Section~\ref{CoordinateDomainTracking}.
On the other hand, the B-Domain tracking aims to maximize the posterior probability of multipath AoA estimates, utilizing C-Domain state prediction and CKM prior information, as will be detailed in Section~\ref{BeamDomainTracking}.
The predictive beamforming is designed to enhance the multipath beam tracking accuracy in the B-Domain, as will be detailed in Section~\ref{predictivebeamform}, which in turn indirectly improves the target tracking accuracy in the C-Domain through the information exchange assisted by the CKM.
The CKM acts as a bridge to enable reciprocal information exchange between the two domains, providing priors for target tracking in the C-Domain via the mapping from multipath channel parameters to potential target positions, and for beam tracking in the B-Domain via the corresponding inverse mapping.

In the C-Domain, the measurement models are addressed separately for two scenarios: LoS-present and LoS-absent conditions. 
Adopt the index $r=1,2$ to denote the LoS-present and LoS-absent cases, respectively.
For LoS-present cases, the C-Domain measurement, vector $\boldsymbol{\beta}_1[\eta] = [\tau_1[\eta], \nu_1[\eta], \cos(\theta_1[\eta])]^T \in \mathbb{C}^{3\times 1}$, comprises the delay, Doppler shift, and the cosine of the AoA for the LoS path.
For LoS-absent cases, vector $\boldsymbol{\beta}_2[\eta]=[\boldsymbol{\tau}[\eta];\boldsymbol{\mu}[\eta];\cos(\boldsymbol{\theta}[\eta])]\in \mathbb{C}^{3P \times 1}$ is defined by the B-Domain estimated path parameters ${\boldsymbol{\tau}}[\eta]$, ${\boldsymbol{\nu}}[\eta]$, and ${\boldsymbol{\theta}}[\eta]$ for all $P$ paths. 
On the other hand, the C-Domain states are defined identically for both LoS-present and LoS-absent cases, which consists of the position and velocity of the moving vehicle, i.e., $\boldsymbol{\alpha}[\eta]=\left[\boldsymbol{q}[\eta],v_1[\eta]\right]^T\in\mathbb{R}^{3\times 1}$.
A measurement model exploiting the geometric relationship that connects $\boldsymbol{\alpha}[\eta]$ and $\boldsymbol{\beta}_1[\eta]$ is employed in the presence of the LoS path; otherwise, the CKM that connecting multipath parameter estimates and the state of a moving vehicle is utilized to construct the measurement model, as will be detailed in Section \ref{coordinate_domian_tracking} of this paper.
Moreover, an identical C-Domain state evolution model, i.e., a constant velocity movement model, is adopted for both LoS-present and LoS-absent cases.
In the $[\eta]$-th time slot, by leveraging the predicted vehicle state $\boldsymbol{\hat{\alpha}}[\eta|\eta-1]$ and the C-Domain measurements $\boldsymbol{\beta}_r[\eta],r=(1,2)$, the RSU applies an EKF-based method to update the state of the vehicle, as will be detailed in Section \ref{coordinate_domian_tracking} of this paper.
Then, based on the C-Domain state evolution model, the state of the vehicle is predicted for the $[\eta+1]$-th slot, i.e., $\boldsymbol{\hat{\alpha}}[\eta+1|\eta]$, which is further exploited for state update in the $[\eta+1]$-th time slot.

In the B-Domain, measurement is the received echo signal at the RSU and state corresponds to the multipath AoA.
At the beginning of time slot $[\eta]$, the RSU first receives echo signals $\boldsymbol{R}[\eta]$ based on the predictive BF matrix $\boldsymbol{F}[\eta|\eta-1]$, facilitating the extraction of the round trip delay $\boldsymbol{\tau}[\eta]\in \mathbb{R}^{P \times 1}$ and Doppler shift $\boldsymbol{\mu}[\eta]\in \mathbb{R}^{P \times 1}$ through the matched filtering technique \cite{sanantonioMIMORadarAmbiguity2007,friedlanderTransmitBeamformingMIMO2012}. 
Note that different propagation paths possess different $\tau_i[\eta]$ and $\nu_i[\eta]$, and thereby the RSU is capable of separating the signals $\boldsymbol{R}_i[\eta]$ reflected from different scatterers.
For the $i$-th path, the state $\{\theta_i[\eta]\}_{i=1}^{P}$ in the B-Domain can be updated by incorporating the state prediction $\{\boldsymbol{p}(\theta_i[\eta|\eta-1])\}_{i=1}^{P}$ from the previous slot and the measurement $\boldsymbol{R}_i[\eta]$ in the current slot, as detailed in Section \ref{beamdomain} of this paper.
The B-Domain state evolution model, which is characterized by the beam TPM, needs to be established for B-Domain tracking.
This model relies on the channel's temporal correlation as well as the \textit{a prior} information stored in the CKM, where the latter one further depends on the predicted state of the vehicle $\hat{\boldsymbol{\alpha}}[\eta+1|\eta]$.
Specifically, in Fig. \ref{fig3}, the matrix $\boldsymbol{\Pi}_i^{\uppercase\expandafter{\romannumeral1}}[\eta]$ denotes the TPM for the $i$-th path due to vehicle movement and the CKM generates the beam TPMs $\boldsymbol{\Pi}_i^{\uppercase\expandafter{\romannumeral2}}[\eta]$ for each path based on the predicted position $\boldsymbol{\hat{q}}[\eta+1|\eta]$ in the C-Domain. 
Consequently, the variation in AoA associated with each path can be effectively modeled as a discrete Markov process \cite{zhangFastBeamTracking2019}. 

Based on the B-Domain state evolution model and the updated B-Domain state $\boldsymbol{\hat{\theta}}[\eta]$, B-Domain state prediction provides two kinds of angle prior information, i.e., the soft prediction $\{\boldsymbol{p}(\theta_i[\eta+1|\eta])\}_{i=1}^{P}$ and the hard prediction $\boldsymbol{\hat{\theta}}[\eta+1|\eta]$, where the former is exploited to assist B-Domain state updates while the latter is employed for designing the predictive beamformer. 
Then, based on the hard prediction $\boldsymbol{\hat{\theta}}[\eta+1|\eta]$, the predictive BF matrix $\boldsymbol{F}[\eta+1|\eta]$ for the next time slot is designed to minimize the maximum AoA estimation error among multiple propagation paths, which determines the received echo signal $\boldsymbol{R}[\eta+1]$ for the next time slot.
By incorporating both the soft prediction $\boldsymbol{p}(\theta_i[\eta+1|\eta])$ and new measurements $\boldsymbol{R}_i[\eta+1]$ in the $[\eta+1]$-th time slot, the B-Domain state $\boldsymbol{\hat{\theta}}[\eta+1]$ can be effectively updated.

Moreover, CKM directly links the coordinate and beam domains and embeds environmental awareness into the tracking process across both domains. 
From the C-Domain to the B-Domain, CKM enhances beam tracking performance via mapping the state prediction of the vehicle to \textit{a prior} AoA information, particularly beneficial for NLoS paths characterized by lower sensing signal-to-noise ratios (SNRs). 
From the B-Domain to the C-Domain, CKM is reversely utilized to infer the state of the vehicle based on multipath parameter measurements for the LoS-absent case.
As a result, continuous C-Domain tracking is maintained for both LoS-present and LoS-absent cases.
Moreover, since the coherence time of the C-Domain state evolution model significantly exceeds that of the B-Domain state evolution model, C-Domain tracking further improves the continuity in B-Domain tracking via exploiting CKM, particularly scenarios involving random path birth and death, or burst change of multipath AoA.
However, the uncertainty and inaccuracy of the CKM affect the performance of the proposed dual-domain tracking. 
In the C-Domain, where the CKM serves as the measurement model for LoS-absent cases, these imperfections directly degrade the accuracy of target state estimation. 
In the B-Domain, where CKM provides partial prior information for multipath AoA estimation, these imperfections indirectly degrade the accuracy of beam tracking.

It is noteworthy that multipath AoA parameter $\boldsymbol{\theta}$ plays different roles in the beam and coordinate domains. 
In the B-Domain, it serves as a state, updated by combining the echo signal $\boldsymbol{R}[\eta]$ with the soft beam state prediction $\left\{\boldsymbol{p}(\theta_i[\eta|\eta-1])\right\}_{i=1}^{P}$. 
It is expected that this prior information $\{\boldsymbol{p}(\theta_i[\eta|\eta-1])\}_{i=1}^{P}$ can enhance the beam state update accuracy, particularly in low SNR scenarios. 
In the C-Domain, $\boldsymbol{\theta}$ is treated as a measurement rather than a state, which is utilized for updating the state of the vehicle. 

In one time slot, the proposed dual-domain tracking scheme alternately updates the states defined in both domains and facilitates information exchanges between them via the CKM.
Across time slots, the proposed dual-domain still follows the traditional Kalman filter framework, where state prediction is performed in both domains to provide \textit{a prior} information for state updates in the next time slot.
Different from traditional passive tracking approaches in the field of radar, ISAC provides additional degrees of freedom for designing sensing signals for the next time slot based on the predicted AoAs of multiple paths.
Consequently, a predictive BF design based on the B-Domain state prediction scheme is critical for improving the performance of the proposed dual-domain tracking.

\textit{Remark 1: }It is worth noting that the proposed dual-domain framework can be naturally extended to the joint tracking of other channel parameters, such as delay and Doppler shifts.
Specifically, the B-Domain state vector can be extended to include the delay and Doppler shift of each path in addition to AoA.
The temporal state evolution can then be modeled by constructing TPMs analogous to those developed for the AoA parameter, and a higher-dimensional CKM that encodes the joint distribution of multiple channel parameters needs to be established.
However, this extension comes at the cost of increased computational complexity in the B-Domain tracking.

\section{CKM-assisted C-Domain Tracking}
\label{coordinate_domian_tracking}
In this section, we propose a CKM-assisted coordinated domain EKF tracking scheme that is applicable for both LoS-present and LoS-absent cases.
The proposed C-Domain tracking scheme requires identifying whether the LoS is present or not, a problem already studied in \cite{zhaoSensingAssistedPredictiveBeamforming} and \cite{zengCKMAssistedIdentificationPredictive2023a}.
\subsection{C-Domain State Evolution Model}
Considering a constant velocity movement model for the moving vehicle within the maximum time of interest \cite{liuRadarAssistedPredictiveBeamforming2020a}, the C-Domain state evolution model is given by $\boldsymbol{\alpha}[\eta+1]=\boldsymbol{E} \boldsymbol{\alpha} [\eta]+\boldsymbol{w}_{\alpha}[\eta+1]$, where $\boldsymbol{E} =\left [ 1,0,\Delta T;0,1,0;0,0,1 \right ]^T\in \mathbb{R}^{3\times 3}$ is the state transition matrix and $\boldsymbol{w}_{\alpha}[\eta+1]= [ w_{q_{x}}[\eta+1], w_{q_{y}}[\eta+1], w_{v}[\eta+1]]^T $ denotes the state evolution noise.
Assuming mutually independent state evolution noises \cite{Wei2023UAV,liuRadarAssistedPredictiveBeamforming2020a}, we have $\boldsymbol{w}_{\alpha }[\eta+1]\sim \mathcal{N} (\boldsymbol{0},\boldsymbol{Q}_{\alpha })$ with $\boldsymbol{Q}_{\alpha }= \operatorname{diag} \{ \sigma ^2_{q_{x}},\sigma ^2_{q_{y}},\sigma ^2_{v} \} $, which can be acquired by the RSU through long-term measurements.

\subsection{C-Domain Measurement Model for LoS-present Case}
\label{Los}
If the LoS path exists in the $[\eta]$-th slot, the C-Domain measurements consist of the AoA, delay, and Doppler shift of the LoS path, which is denoted with the subscript $''1''$. Based on the geometric relationships depicted in Fig. \ref{figsystemmodel}, the C-Domain measurement model is given by
\begin{equation}
\label{measurement_model}
    \begin{cases}
  & \tau_1 [\eta ]=\frac{2||\boldsymbol{q}[\eta ]||_2}{c} +w_{\tau}[\eta]\in \mathbb{R}, \\
  & \mu_1 [\eta ]=- \frac{2 v_1[\eta ] q_y[\eta] f_c}{c||\boldsymbol{q}[\eta ]||_2} +w_{\mu}[\eta]\in \mathbb{R}, \\
  & \cos (\theta _1[\eta])=\frac{q_{x}[\eta ]}{||\boldsymbol{q}[\eta ]||_2} +w_{\cos \theta }[\eta] \in (-1,1),
\end{cases}
\end{equation}
where $c$ is the speed of light and $f_c$ is the signal carrier frequency. 
Gaussian random variables $w_{\tau}[\eta]$, $w_{\mu}[\eta]$, and $w_{\cos \theta}[\eta]$ denote the corresponding measurement noises with zero mean and variances $\sigma^2_{\tau}$, $\sigma^2_{\mu}$, and $\sigma^2_{\cos \theta}$, respectively. 
Collecting all the C-Domain observable parameters in a vector $ \boldsymbol{\beta}_1[\eta]=[\tau_1[\eta],\mu_1[\eta],\cos(\theta _1[\eta])]^T\in\mathbb{R}^{3\times 1}$, the measurement model is given by
\begin{equation}
\label{LosModel}
\boldsymbol{\beta}_1[\eta]=g_{1}(\boldsymbol{\alpha}[\eta])+\boldsymbol{w}_{\beta_1}[\eta],
\end{equation} 
where $\boldsymbol{w}_{\beta_1}[\eta]= [w_{\tau}[\eta], w_{\mu}[\eta],w_{\cos \theta}[\eta]]^T \sim \mathcal{N}(\boldsymbol{0},\boldsymbol{Q}_{\beta_1})$ and $\boldsymbol{Q}_{\beta_1}= \operatorname{diag} \{ \sigma ^2_{\tau},\sigma ^2_{\mu},\sigma ^2_{\cos \theta }\}$. 
The non-linear measurement function $g_{1}(\cdot): \mathbb{R}^{3\times 1}\to \mathbb{R}^{3\times 1}$ is defined by (\ref{measurement_model}).

\subsection{C-Domain Measurement Model for LoS-absent Case}
\label{NLoS}
In practice, the LoS path might be shadowed by surrounding buildings or large scatterers in neighboring lanes. 
In LoS-absent cases, the C-Domain measurement vector is given by
$\boldsymbol{\beta}_2[\eta] =[\boldsymbol{\tau}[\eta];\boldsymbol{\mu}[\eta];\cos(\boldsymbol{\theta}[\eta])]\in \mathbb{C}^{3P\times 1}$, where $\boldsymbol{\tau}[\eta]=\left[\tau_1[\eta],\dots,\tau_P[\eta]\right]^T$, $\boldsymbol{\mu}[\eta]=\left[\mu_1[\eta],\dots,\mu_P[\eta]\right]^T$, and $\cos(\boldsymbol{\theta}[\eta])=\left[\cos(\theta_1[\eta]),\dots,\cos(\theta_P[\eta])\right]^T$. 
For LoS-absent cases, the measured NLoS path parameters are not directly related to the state of the vehicle. 
Fortunately, CKM provides the location-specific channel parameters, which maps the position and the velocity of the vehicle to the path delays, Doppler shifts, and angular information.
Therefore, the CKM itself provides a measurement function which is specified as $g_{2}(\cdot): \mathbb{R}^{3\times1}\to \mathbb{R}^{3{P} \times 1}$ that maps the state vector \(\boldsymbol{\alpha}[\eta]\) to the measurement vector $\boldsymbol{\beta}_2[\eta]$.
In other words, the measurement model is given by
\begin{equation}
\label{NLoSModel}
\boldsymbol{\beta}_2[\eta] = g_{2}(\boldsymbol{\alpha}[\eta]) + \boldsymbol{w}_{\beta_2}[\eta],
\end{equation}
where \(\boldsymbol{w}_{\beta_2 }[\eta]\sim \mathcal{N}(0,\boldsymbol{Q}_{\beta_2})\in \mathbb{R}^{3{P}\times 1} \) is the measurement noise vector and the corresponding measurement noise covariance matrix for the LoS-absent case is assumed\footnote{In this paper, we assume the same measurement noise variance for the same type of channel parameter estimation in \eqref{LosModel} and \eqref{NLoSModel} as the channel parameters stored in the CKM come from long-term channel measurement as well \cite{zengTutorialEnvironmentAwareCommunications2023}.} as $\boldsymbol{Q}_{\beta_2}=\operatorname{diag}  \{ \sigma_\tau^2 \boldsymbol{I}_P,\sigma_\mu^2 \boldsymbol{I}_P,\sigma_{\cos \theta}^2 \boldsymbol{I}_P \}$.
As commonly adopted in literature \cite{liuRadarAssistedPredictiveBeamforming2020a,zengCKMAssistedIdentificationPredictive2023a}, we assume the estimation errors of C-Domain measurement to be Gaussian distributed, as required by the EKF framework. 
In practice, $\sigma_\tau^2$, $\sigma_{\cos \theta}^2$, and $\sigma_{\cos \theta}^2$ depend on antenna aperture, bandwidth, frame length, and even BF strategy. 
The relationship between path parameter estimation errors and those resources above is specifically determined by parameter estimation methods. 
Therefore, for simplicity, this paper assumes fixed measurement noise variances in the adopted C-Domain.

\subsection{EKF-based C-Domain Tracking}
\label{CoordinateDomainTracking}
Due to the non-linear measurement model for both the LoS-present case in (\ref{LosModel}) and LoS-absent case in (\ref{NLoSModel}), we adopt the EKF \cite{kay1993fundamentals} to track the state of the vehicle in the C-Domain. 
In the $\eta$-slot, assuming we have the knowledge of the state vector ${\boldsymbol{\alpha}}[\eta]\in\mathbb{R}^{3\times 1}$ estimated at the $\eta$-slot, the state prediction for the $[\eta+1]$-slot is given by
\begin{equation}
\label{stateprediction}
       \hat{\boldsymbol{\alpha}}[\eta+1|\eta]=\boldsymbol{E}{\boldsymbol{\alpha}}[\eta],
\end{equation}
and the prediction covariance matrix $\boldsymbol{C}_{r}[\eta+1|\eta]\in\mathbb{R}^{3\times3}$, $r=1,2$ is given by $\boldsymbol{C}_{r}[\eta+1|\eta]=\boldsymbol{E} \boldsymbol{C}_{r}[\eta] \boldsymbol{E}^T + \boldsymbol{Q}_{\alpha}$, where $\boldsymbol{C}_{r}[\eta]$ is the posterior covariance matrix for the C-Domain state update in the $[\eta]$-th slot.

\subsubsection{LoS-present case}
To update the state in the $[\eta]$-th time slot, the EKF employs a linearized measurement model around the predicted state\footnote{In our notation, only the state prediction and state update are denoted with a hat (e.g., $\hat{\boldsymbol{\alpha}}[\eta|\eta-1]$ and $\hat{\boldsymbol{\alpha}}[\eta]$); other variables, such as the measurement $\boldsymbol{\beta}[\eta|\eta-1]$ and $\boldsymbol{\beta}[\eta]$, are presented without a hat.}.
Specifically, we have: $\boldsymbol{\beta}_1[\eta ]\approx g_{1}(\hat{\boldsymbol{\alpha}}[\eta |\eta -1])+\boldsymbol{G}_1(\boldsymbol{\alpha}[\eta ]-\hat{\boldsymbol{\alpha}}[\eta|\eta -1])+\boldsymbol{w}_{\beta_1}[\eta]$, where $\hat{\boldsymbol{\alpha}}[\eta |\eta -1]$ is the predicted C-Domain state in $[\eta-1]$-th time slot, $\boldsymbol{G}_1\in \mathbb{R}^{3\times 3}$ represents the Jacobian matrix of $g_{1}$ with respect to the state vector and it is given by (\ref{G_1}) 
\begin{equation}
\small
\label{G_1}
\begin{aligned}
\boldsymbol{G}_1 & =\left.\frac{\partial {g}_1}{\partial \boldsymbol{\alpha }}\right|_{\hat{\boldsymbol{\alpha}}[\eta |\eta -1]} 
= \left. \begin{bmatrix}
\frac{\partial \tau_1[\eta]}{\partial q_{x}[\eta ]}   & \frac{\partial \tau_1[\eta]}{\partial q_{y}[\eta ]} & 0 \\
 \frac{\partial \mu_1[\eta]}{\partial q_{x}[\eta ]} & \frac{\partial \mu_1[\eta]}{\partial q_{y}[\eta ]} & \frac{\partial \mu_1[\eta]}{\partial v_{1}[\eta ]} \\
 \frac{\partial  \cos(\theta _1[\eta])}{\partial q_{x}[\eta ]} &  \frac{\partial  \cos(\theta _1[\eta])}{\partial q_{y}[\eta ]} & 0 
\end{bmatrix}\right|_{\hat{\boldsymbol{\alpha}}[\eta |\eta -1]}\\ 
&=\left.\begin{bmatrix}
 \frac{2 q_x[\eta ]}{c ||\boldsymbol{q}[\eta ]||_2}  & \frac{2 q_y[\eta ]}{c ||\boldsymbol{q}[\eta ]||_2}  &0 \\
 -\frac{2 v_1[\eta ]f_c q_y[\eta ] q_x[\eta]}{c||\boldsymbol{q}[\eta ]||_2^{3 }}  & \frac{2 f_c v_1[\eta ](q_x[\eta ])^2}{c||\boldsymbol{q}[\eta ]||_2^{3 }} & -\frac{2 q_y[\eta ]f_c }{c||\boldsymbol{q}[\eta ]||_2} \\
 \frac{q_y[\eta ]^2}{||\boldsymbol{q}[\eta ]||_2^{3 }}  &  -\frac{q_x[\eta ] q_y[\eta ]}{||\boldsymbol{q}[\eta ]||_2^{3 }}  & 0
\end{bmatrix}\right|_{\hat{\boldsymbol{\alpha}}[\eta |\eta -1]} .
\end{aligned}
\end{equation}
By substituting the state prediction $\hat{\boldsymbol{\alpha}}[\eta|\eta-1]$ into the equation above, $\boldsymbol{G}_1$ is obtained. 

\subsubsection{LoS-absent case}
In the LoS-absent case, the measurement function $g_{2}(\cdot)$ in (\ref{NLoSModel}) leveraging CKM provides a look-up table for obtaining multipath parameter measurements based on the location of the vehicle.
To update the state of the vehicle, a linearized measurement model can be established:
\begin{equation}
{\boldsymbol{\beta}_2}[\eta] \approx g_{2}(\hat{\boldsymbol{\alpha}}[\eta|\eta-1]) + \boldsymbol{G}_2(\boldsymbol{\alpha}[\eta] - \hat{\boldsymbol{\alpha}}[\eta|\eta-1]) + \boldsymbol{w}_{\beta_2}[\eta],
\end{equation}
where $g_{2}(\hat{\boldsymbol{\alpha}}[\eta|\eta-1])\in \mathbb{R}^{3P \times 1}$ can be obtained via referring to CKM based on the predicted vehicle state.
In the LoS-absent case, the AoA of NLoS paths can vary significantly with slight vehicle movement due to multi-path propagation. 
Consequently, a high spatial resolution CKM is essential for capturing such angular variations.
The Jacobian matrix \(\boldsymbol{G}_2\) represents the partial derivatives of the measurement function \(g_{2}(\cdot)\) with respect to the C-Domain state vector and is computed numerically as follow: 
\begin{equation}
\label{G_2}
\begin{aligned}
    \boldsymbol{G}_2
&=\left.  \frac{\partial g_{2}}{\partial \boldsymbol{\alpha }} \right|_{\hat{\boldsymbol{\alpha}}[\eta |\eta -1]} 
=\left.  [\frac{\partial g_{2}}{\partial q_x} ,\frac{\partial g_{2}}{\partial q_y},\frac{\partial g_{2}}{\partial v_1}  ]\right|_{\hat{\boldsymbol{\alpha}}[\eta |\eta -1]},\\
\end{aligned}
\end{equation}
where $\frac{\partial g_{2}}{\partial q_x}= {[g_{2}(q_x[\eta ]+\Delta q_x,q_y[\eta ],v_1[\eta ])-g_{2}(\boldsymbol{\alpha}[\eta ])]}/{\Delta q_x}$, $\frac{\partial g_{2}}{\partial q_y}={[(g_{2}(q_x[\eta ],q_y[\eta ]+\Delta q_y,v_1[\eta ])-g_{2}(\boldsymbol{\alpha}[\eta ])]}/{\Delta q_y} $, and $\frac{\partial g_{2}}{\partial v_1}={[(g_{2}(q_x[\eta ],q_y[\eta ],v_1[\eta ]+\Delta v_1)-g_{2}(\boldsymbol{\alpha}[\eta ])]}/{\Delta v_1}$, $\Delta q_x$, $\Delta q_y$, and $\Delta v_1$ denote small perturbations in the C-Domain state.
Note that each element of $\boldsymbol{G}_2$ quantifies how the multipath parameters change with respect to the C-Domain state vector.

Following the standard EKF framework \cite{liuRadarAssistedPredictiveBeamforming2020a}, the procedures for proposed C-Domain tracking in both LoS-present $(r=1)$ and LoS-absent $(r=2)$ scenarios are outlined below:
(1) Linearization: $ \boldsymbol{\beta}_r[\eta ]\approx g_{r}(\hat{\boldsymbol{\alpha}}[\eta |\eta -1])+\boldsymbol{G}_r(\boldsymbol{\alpha}[\eta ]-\hat{\boldsymbol{\alpha}}[\eta|\eta -1])+\boldsymbol{w}_{\beta_r}[\eta ] $, 
(2) Kalman gain matrix: $\boldsymbol{K}_r=\boldsymbol{C}_{r}[\eta|\eta-1] \boldsymbol{G}_r^T(\boldsymbol{G}_r \boldsymbol{C}_{r}[\eta|\eta-1]\boldsymbol{G}_r^T+\boldsymbol{Q}_{\beta_r})^{-1}$, 
(3) State update: $\hat{\boldsymbol{\alpha}}[\eta]=\hat{\boldsymbol{\alpha}}[\eta|\eta-1]+\boldsymbol{K}_r(\boldsymbol{\beta}_r[\eta]-g_{r}(\hat{\boldsymbol{\alpha}}[\eta|\eta-1]))$, 
(4) Posterior covariance matrix: $\boldsymbol{C}_{r}[\eta]=(\boldsymbol{I}_3-\boldsymbol{K}_r \boldsymbol{G}_r)\boldsymbol{C}_{r}[\eta|\eta-1]$, 
(5) State prediction: $\hat{\boldsymbol{\alpha}}[\eta+1|\eta]=\boldsymbol{E} \hat{\boldsymbol{\alpha}}[\eta]$, 
(6) Prediction covariance matrix: $\boldsymbol{C}_{r}[\eta+1|\eta]=\boldsymbol{E} \boldsymbol{C}_{r}[\eta] \boldsymbol{E}^T + \boldsymbol{Q}_{\alpha}$.
Interested readers can refer to \cite{liuRadarAssistedPredictiveBeamforming2020a,zengCKMAssistedIdentificationPredictive2023a,Wu2023IOT,Wei2023UAV} for further details of EKF.
Note that $\boldsymbol{\beta}_r[\eta]$ is always obtained from the received signals. 
In the LoS-absent case, the measurement model is characterized by the CKM, which provides a numerical mapping from the target state to the multipath parameters, thereby ensuring continuous and stable state update.
The computational complexity of the proposed EKF scheme is constant, i.e., $\mathcal{O}(1)$, in the LoS-present case, whereas in the LoS-absent case, it increases to $\mathcal{O}(P^3)$ due to the inversion of a $3P \times 3P$ matrix required in the Kalman gain computation.

\section{CKM-assisted B-Domain Tracking}
\label{beamdomain}

\subsection{B-Domain State Evolution Model}
Similar to \cite{zhangTrainingBeamSequence2022a} and \cite{zhangFastBeamTracking2019}, define $\Theta\triangleq\{\tilde{\theta}_\kappa={\pi(\kappa-1)}/{N_\theta}\}^{N_\theta}_{\kappa=1}$ as an angular grid with ${N}_\theta$ equal segments spanning the entire angular space $[0, \pi)$.
Also, the B-Domain state evolution model is defined as a discrete Markov process, i.e.,
\begin{equation}
\label{pPi}
    \boldsymbol{p}({\theta}_i[\eta+1|\eta] )=\boldsymbol{p}(\theta _i[\eta])\boldsymbol{\Pi}_i[\eta]\in[0,1]^{1\times N_{\theta }},
\end{equation}
where $\boldsymbol{\Pi}_i [\eta]\in \mathbb{R}^{N_\theta \times N_\theta}$ denotes the TPM for the AoA of $i$-th path in the $[\eta]$-th time slot, $\boldsymbol{p}({\theta }_i[\eta])$ and $\boldsymbol{p}({\theta}_i[\eta+1|\eta])$ denote the probability mass function (PMF) of the AoA for the $i$-th path in the $[\eta]$-th slot and the predicted PMF of the AoA of $i$-th path in the $[\eta+1]$-th time slot, respectively.
The $(\kappa,\iota)$-th element of the TPM for the $i$-th path in the $[\eta]$-th slot is given by $\boldsymbol{\Pi}_i [\eta]_{\kappa,\iota}=\operatorname{Pr}\{ \theta_i[\eta+1]= \tilde{\theta}_{\iota} \mid \theta_i[\eta]= \tilde{\theta}_{\kappa}\}, \forall \kappa,\iota\in \{ 1,\dots, N_\theta \}$, 
where $\operatorname{Pr}\{ \theta_i[\eta+1]= \tilde{\theta}_{\iota} \mid \theta_i[\eta]= \tilde{\theta}_{\kappa}\}$ denotes the transition probability of the AoA of the $i$-th path from angle $\tilde{\theta}_{\kappa}$ at time slot $[\eta]$ to $\tilde{\theta}_{\iota}$ at time slot $[\eta+1]$. 
Clearly, the TPM should satisfy the following requirements:
(1) ${0} \le \boldsymbol{\Pi}_i [\eta]_{\kappa,\iota} \le {1}$ for $\forall \kappa,\iota\in \{ 1,\dots, N_\theta \}$.
(2) The sum of each row of $\boldsymbol{\Pi}_i [\eta]
$ should be equal to 1, i.e., $|| \boldsymbol{\Pi}_i [\eta]_{\kappa,:}||_1=1, \forall \kappa \in \{ 1,\dots, N_\theta \}$.

Both the channel temporal correlation and the CKM provide \textit{a prior} information on beam transition and thus affect the TPM.
We model the fusion of these two sources as a convex combination of TPMs and such a convex combination remains mathematically valid, as established in \cite{marcus1992survey,Baake_Sumner_2022}.
Therefore, the angular TPM for the $i$-th path at $[\eta]$-th slot is modeled as: 
\begin{equation}
\label{Pi}
\boldsymbol{\Pi}_i[\eta]=(1-c_{\Pi_i}[\eta]) \cdot \boldsymbol{\Pi}_i^{\uppercase\expandafter{\romannumeral1}}[\eta]  
+ c_{\Pi_i}[\eta] \cdot \boldsymbol{\Pi}_i^{\uppercase\expandafter{\romannumeral2}}[\eta],
\end{equation}
where $\boldsymbol{\Pi}_i^{\uppercase\expandafter{\romannumeral1}}[\eta]$ denotes the TPM caused by the channel temporal correlation and $\boldsymbol{\Pi}_i^{\uppercase\expandafter{\romannumeral2}}[\eta]$ denotes the TPM designed using the channel \textit{a priori} information provided by CKM.
The real constant $c_{\Pi_i}[\eta] \in [0,1]$ is the probability of selecting the transition mechanism $\boldsymbol{\Pi}_i^{\uppercase\expandafter{\romannumeral2}}[\eta]$ provided by CKM, while $(1-c_{\Pi_i}[\eta])$ corresponds to the probability of selecting the transition mechanism $\boldsymbol{\Pi}_i^{\uppercase\expandafter{\romannumeral1}}[\eta]$ provided by channel temporal correlation.
The value of $c_{\Pi_i}[\eta]$ depends on the condition of beam transition, the received echo SNR, the horizontal speed of the target vehicle relative to the RSU, and the quality of the constructed CKM.

\begin{figure}[!t]
\centering
\includegraphics[width=3.3 in]{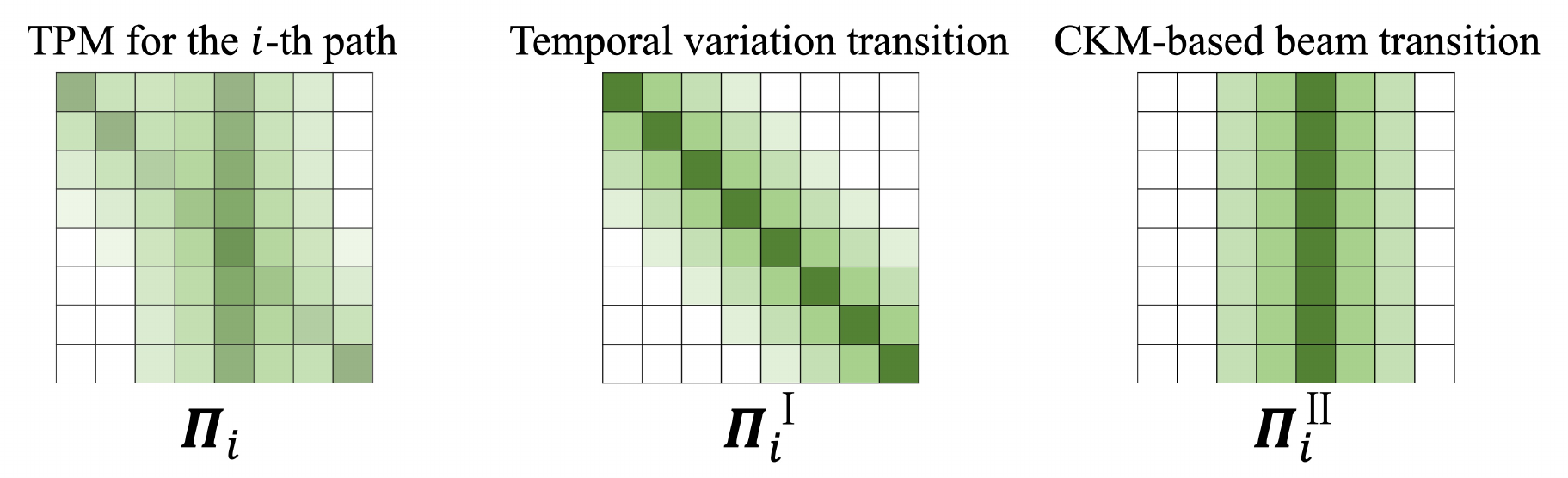}
\caption{Angular TPM for the $i$-th path.}
\label{fig_TPM}
\end{figure}

As mentioned earlier, the AoA of the $i$-th path between two consecutive time slots should remain close even in high-mobility wireless networks, provided there is no random birth/death of paths.
Following \cite{zhangTrainingBeamSequence2022a} and \cite{chenBeamTrainingTracking}, the $(\kappa,\iota)$-th element of \(\boldsymbol{\Pi}_{i}^{\uppercase\expandafter{\romannumeral1}}\) of a path in the $[\eta]$-th slot is given by 
\begin{equation}
\label{Pi_2}
\left \{ \boldsymbol{\Pi}_i^{\uppercase\expandafter{\romannumeral1}} [\eta]\right \}_{\kappa,\iota}=
\begin{cases}
    \zeta \xi^{\left|\kappa-\iota\right|},&{\text{if}}\left|\kappa-\iota\right| \leq \epsilon ,\\
    {0}, &{\text{otherwise,}}
\end{cases}
\end{equation}
where $\forall i\in \{1,\dots,P\}$. 
The parameter \(\xi \in [0,1]\) characterizes the variation speed of angles over time, where \(\xi \rightarrow 0\) indicates almost no change and \(\xi \rightarrow 1\) indicates significant change. 
The variable \(\epsilon\) represents the boundary of AoA variation and \(\zeta\) is a normalization coefficient as in \cite{seoTrainingBeamSequence2016a,zhangFastBeamTracking2019}. 
Both \(\xi\) and \(\epsilon\) depend on the relative speed of the vehicle to the RSU, given by $\sin (\hat{\theta} _1[\eta+1|\eta]) \hat{v}_1[\eta+1|\eta]$. 
On the other hand, the $(\kappa,\iota)$-th element of \(\boldsymbol{\Pi}_{i}^{\uppercase\expandafter{\romannumeral2}}\) in the $[\eta]$-th slot is given by 
\begin{equation}
\label{Pi_1}
\left \{ \boldsymbol{\Pi}_i^{\uppercase\expandafter{\romannumeral2}}[\eta] \right \}_{\kappa,\iota}= \varsigma \exp \left (- \frac{(\hat{\theta}_i[\eta+1|\eta]-\tilde {\theta_\iota} )^2}{2 \sigma_\text{CKM}^2}  \right )   , \forall \kappa,
\end{equation}
where $\varsigma=1/\left(\sum_{\iota =1}^{N_\theta}  \exp\left (- \frac{(\hat{\theta}_i[\eta+1|\eta]-\tilde {\theta_\iota} )^2}{2 \sigma_\text{CKM}^2}  \right ) \right)$ is the normalization coefficient, $\hat{\theta}_i[\eta+1|\eta]$ is the predicted AoA of the $i$-th path provided by CKM in (\ref{CKM}) based on the predicted vehicle state $\hat{\boldsymbol{\alpha}}[\eta+1|\eta]$ in (\ref{stateprediction}).
$\tilde{\theta}_\iota=\pi(\iota -1)/ N_\theta$ denotes the $\iota$-th angle on the grid $\Theta$ and $\sigma_\text{CKM}^2$ serves as a variance of beam transition accounting for noise in the angle measurements and uncertainty in the CKM.
A typical example of $\boldsymbol{\Pi}_{i}$, $\boldsymbol{\Pi}_{i}^{\uppercase\expandafter{\romannumeral1}}$, and $\boldsymbol{\Pi}_{i}^{\uppercase\expandafter{\romannumeral2}}$ are illustrated in Fig. \ref{fig_TPM}.

Moreover, the B-Domain state evolution model should depend on the condition of beam transition.
\begin{figure}[!t]
\centering
\includegraphics[width=3.1 in]{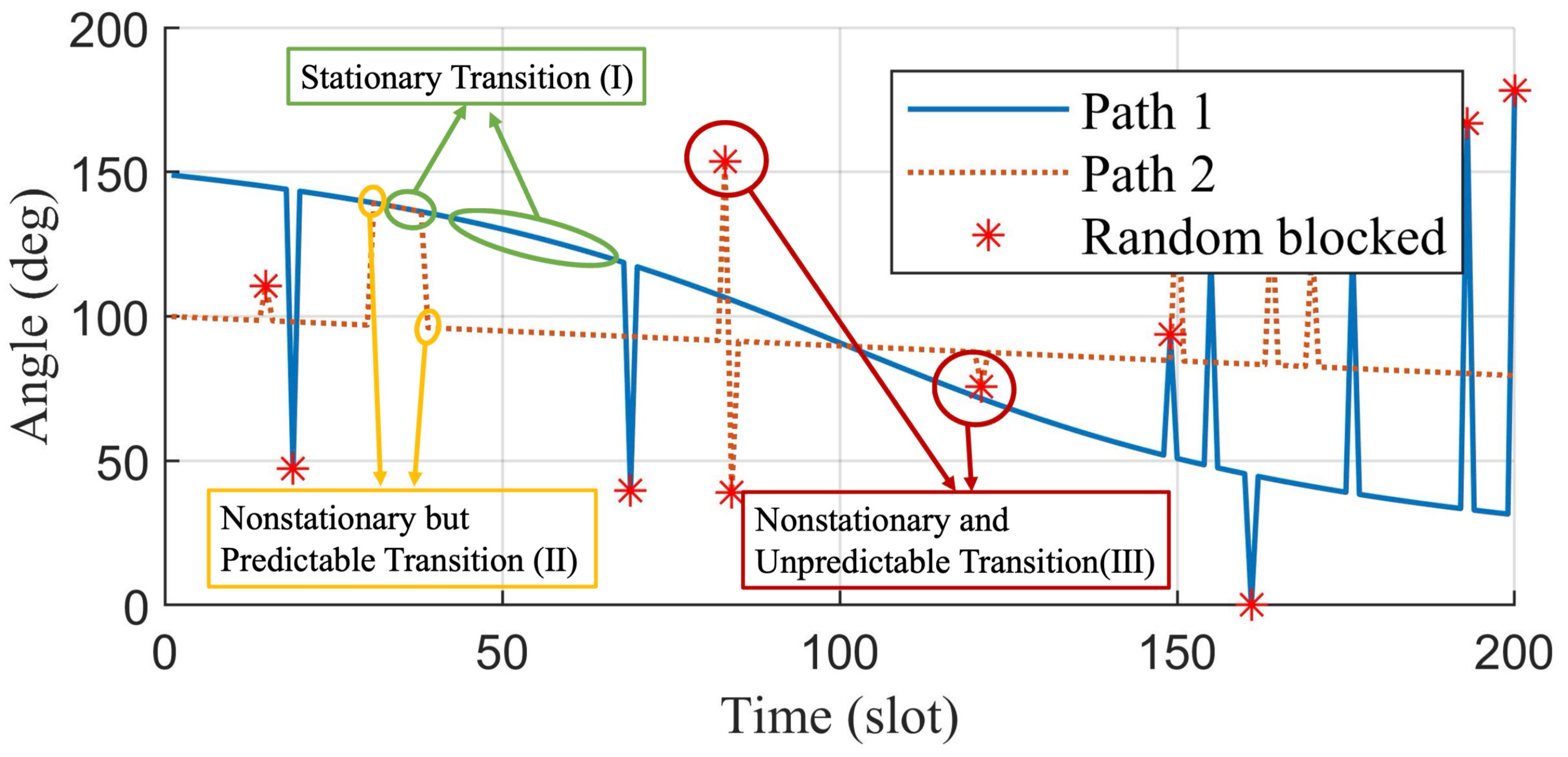}
\caption{Angular variation over time for each path.}
\label{fig_muitpath}
\end{figure}
In Fig.\ref{fig_muitpath}, the angular variations of two paths over time are demonstrated, where \textbf{path 1} represents a LoS path and \textbf{path 2} represents an NLoS path. 
During the time of interest, the condition of beam transition can be established according to the following three types of angular variations:
\begin{itemize}
    \item Stationary Transition $(\uppercase\expandafter{\romannumeral1})$: In this case, both $\boldsymbol{\Pi}_i^{\uppercase\expandafter{\romannumeral1}}$ and $\boldsymbol{\Pi}_i^{\uppercase\expandafter{\romannumeral2}}$ are available due to the time-correlated variations in AoA and the CKM mapping between the position of vehicle and AoA. 
    In this case, $c_{\Pi_i}\in (0,1)$.
    A larger $c_{\Pi_i}$ is preferred under a low received echo SNR or the target vehicle traveling with a high horizontal speed relative to the RSU, where the weaker channel temporal correlation or unreliable measurements necessitate greater reliance on the CKM prior.
    In contrast, a lower quality constructed CKM shifts reliance toward temporal channel correlation, leading to a smaller $c_{\Pi_i}$.

    \item Non-stationary but Predictable Transition $(\uppercase\expandafter{\romannumeral2})$: The angle may abruptly change due to the fixed scatterers in the environment which have been recorded in the CKM (shown by the yellow arrows in the figure). In this case, the angle transition is not time-correlated and $\boldsymbol{\Pi}_i^{\uppercase\expandafter{\romannumeral1}}$ can not be exploited. Fortunately, CKM has stored this beam transition and thus the angle transition is still predictable based on $\boldsymbol{\Pi}_i^{\uppercase\expandafter{\romannumeral2}}$. In this case $c_{\Pi_i}$ is equal to 1.
    \item Non-stationary and Unpredictable Transition $(\uppercase\expandafter{\romannumeral3})$: The random block is caused by randomly appearing scatterers/obstacles in the environment (illustrated by the red asterisks in the figure). For this case, AoA prediction is impossible and no \textit{a priori} information can be used. In this case, a confirm distribution can be adopted, i.e., $ \boldsymbol{p}({\theta}_i[\eta+1|\eta] ) = \frac{1}{N_\theta}\cdot \boldsymbol{1}_{N_\theta}$.
\end{itemize}

The hypothesis testing for LoS path blockage has been proposed in \cite{zhaoSensingAssistedPredictiveBeamforming} and \cite{zengCKMAssistedIdentificationPredictive2023a} for improving target tracking performance.
Similarly, based on the estimated delay and Doppler shift between consecutive time slots, the RSU can determine the condition of beam transition for each path.
In this paper, we assume that the hypothesis testing for cases of the angular variation model is perfect for simplicity, which facilitates the B-Domain state evolution model establishment in the proposed dual-domain tracking and predictive beam forming framework.

\subsection{B-Domain Measurement Model}

After receiving the reflected echo, the RSU can calculate the cross-ambiguity function (CAF) \cite{sanantonioMIMORadarAmbiguity2007} between the received signal $\boldsymbol{R}$ and the transmitted signal $\tilde{\boldsymbol{S}}$ to estimate the delay and Doppler shift of multiple paths.
Following (\ref{R}), a peak occurs at the matched filter output when $\tau$ and $\mu$ are perfectly tuned to the corresponding delay and Doppler shifts of the $i$-th path, respectively, i.e., $\{ \hat{l}_i,\hat{k}_i  \} ={\text{arg max}}_{l_i,k_i} || \boldsymbol{R}\boldsymbol{\Lambda}_{k_i}^H \boldsymbol{S}_{l_i}^H ||_F^2 ,i=1,\dots,P $.
\begin{lemma}
\label{lemma1}
Assuming $L\to \infty $ and $\frac{1}{T_p}\to \infty$, having obtained the perfect delay and Doppler shift estimates, the received signal $\boldsymbol{R}_i$ associated with the $i$-th path can be separated and is given by
\begin{equation}
\label{R_j2}
   \boldsymbol{R}_i \approx L \beta_i \boldsymbol{b}(\theta_i) \boldsymbol{a}^H(\theta_i)\boldsymbol{F} + \boldsymbol{Z}\boldsymbol{\Lambda}_{k_i}^H \boldsymbol{S}_{l_i}^H 
\in \mathbb{C}^{N_r \times N_s},
\end{equation}
for any $i=1,\cdots,P $.
\end{lemma}
\begin{IEEEproof}
Substitute \eqref{R} into the received signal after matched filtering:
\begin{equation}\small
\label{eq_R_j}
\boldsymbol{R}_i
=\underbrace{\sum_{j=1}^P \beta_j \boldsymbol{b}(\theta_j) \boldsymbol{a}^H(\theta_j) \boldsymbol{F} {\boldsymbol{S}_{l_j}}
{\boldsymbol{\Lambda}_{k_j}}\boldsymbol{\Lambda}_{k_i}^H
\boldsymbol{S}_{l_i}^H}_{\text{Signal and interference: } \boldsymbol{R}_i'} +   \underbrace{\boldsymbol{Z}\boldsymbol{\Lambda}_{k_i}^H
\boldsymbol{S}_{l_i}^H}_{\text{Noise: }\boldsymbol{R}_i''}.
\end{equation}
%
To facilitate the derivation, without loss of generality, we assume $l_j\le l_i $, $\forall j< i$. $\boldsymbol{R}_i'$ can be simplified as: 
\begin{equation}
\small
\label{eq_R_j_1}
\begin{aligned}
 \boldsymbol{R}_i'=& \beta_i \boldsymbol{b}(\theta_i) \boldsymbol{a}^H(\theta_i)\boldsymbol{F} \boldsymbol{S}\boldsymbol{S}^H \\ &+ \sum_{j \neq i}^P \beta_j \boldsymbol{b}(\theta_j) \boldsymbol{a}^H(\theta_j) \boldsymbol{F} {\boldsymbol{S}_{l_j}} \operatorname{diag}(\boldsymbol{0}_{l_i},\boldsymbol{\Lambda}_{ji},\boldsymbol{0}_{[\max (l_j) - l_j]})\boldsymbol{S}_{l_i}^H ,
\end{aligned}
\end{equation}
where 
\begin{equation*}
\small
\label{Lambda_ij}
\begin{aligned}
\boldsymbol{\Lambda}_{ji} 
    &\triangleq \{ \boldsymbol{\Lambda}^{\text{shift}}_{k_i}  \}_{ (l_i-l_j+1):L,(l_i-l_j+1):L}   \{ \boldsymbol{\Lambda}^{\text{shift}}_{k_i}   \}_{1:(L-l_i+l_j),1:(L-l_i+l_j)} ^H. 
\end{aligned}
\end{equation*}
It follows that
\begin{equation}
\small
\begin{aligned}
 \boldsymbol{R}_i' =& \beta_i \boldsymbol{b}(\theta_i) \boldsymbol{a}^H(\theta_i)\boldsymbol{F} \boldsymbol{S}\boldsymbol{S}^H\\ &+ \sum_{j \neq i}^P \beta_j \boldsymbol{b}(\theta_j) \boldsymbol{a}^H(\theta_j) \boldsymbol{F} 
\left( \sum_{l = l_i+1}^{L + l_j} \boldsymbol{x}_{L-l_j} \boldsymbol{x}_{L-l_i}^H \left \{ \boldsymbol{\Lambda}_{ji} \right \}_{l,l} \right) \\
\approx& L \beta_i \boldsymbol{b}(\theta_i) \boldsymbol{a}^H(\theta_i)\boldsymbol{F}
\end{aligned}
\end{equation}
Therefore, for any $i=1,\cdots,P $, $\boldsymbol{R}_i$ is obtained by $\boldsymbol{R}_i \approx L \beta_i \boldsymbol{b}(\theta_i) \boldsymbol{a}^H(\theta_i)\boldsymbol{F} + \boldsymbol{Z}\boldsymbol{\Lambda}_{k_i}^H \boldsymbol{S}_{l_i}^H 
\in \mathbb{C}^{N_r \times N_s}$.
\end{IEEEproof}

Lemma 1 enables the RSU to separate the received signal from different paths for AoA estimation using the estimated delay and Doppler shift. 
According to (\ref{R_j2}), the received signal associated with the $i$-th path is rewritten as: $\boldsymbol{r}_i \triangleq {\text{vec}}(\boldsymbol{R}_i) =L \beta_i (\boldsymbol{F}^T \otimes \boldsymbol{b}(\theta_i) ) \boldsymbol{a}^*(\theta_i) + (\boldsymbol{S}_{l_i}^*\otimes \boldsymbol{Z})  {\text{vec}} (\boldsymbol{\Lambda}_{k_i}^H ) 
 \in \mathbb{C}^{N_sN_r \times 1}$.

\begin{lemma}
\label{lemma2}
Given the AoA of the $i$-th path $\theta_i$, $ \boldsymbol{r}_i$  is a complex multivariate Gaussian random variable with $\boldsymbol{r}_i\sim \mathcal{CN}(   L \beta_i (\boldsymbol{F}^T \otimes \boldsymbol{b}(\theta_i) ) \boldsymbol{a}^*(\theta_i),\sigma_z^2 L \boldsymbol{I}_{N_sN_r})$.
\end{lemma}
\begin{IEEEproof}
It is clear that the mean of $\boldsymbol{r}_i$ is $\mathbb{E} \left \{ \boldsymbol{r}_i\right \}
    = L \beta_i (\boldsymbol{F}^T \otimes \boldsymbol{b}(\theta_i) ) \boldsymbol{a}^*(\theta_i)$.
Then, we proceed with the derivation of the covariance matrix for $\boldsymbol{r}_i$. Consider a $N_r$-dimensional vector denoted as $\tilde{\boldsymbol{z}}_n$, defined as $\tilde{\boldsymbol{z}}_n=\boldsymbol{Z}_{:,(l_i+1):(l_i+L)}\boldsymbol{D}^H_{n,:}$, where $\boldsymbol{Z}_{:,(l_i+1):(l_i+L)}\in\mathbb{C}^{N_r\times L}$ and $\boldsymbol{D}=\boldsymbol{S}\boldsymbol{\Lambda}^{\text{shift}}_{k_i} \in\mathbb{C}^{N_s \times L}$. 

Consequently, we have
\begin{equation}
\begin{aligned}
\left\{\tilde{\boldsymbol{z}}_n\right\}_m &=\boldsymbol{Z}_{m,(l_i+1):(l_i+L)}\boldsymbol{D}^H_{n,:} \\
    &=\sum_{l=1}^{L} \boldsymbol{Z}_{m,(l_i+l)} \boldsymbol{S}^*_{n,l}e^{-j2\pi \mu_ilT_p} ,
\end{aligned}
\end{equation}
and its variance is given by $\sigma^2 \left( \left\{\tilde{\boldsymbol{z}}_n\right\}_m \right)=\sigma^2_z\sum_{l=1}^{L}|\boldsymbol{S}^*_{n,l}e^{-j2\pi \mu_ilT_p}|^2=\sigma^2_z L$.
Given that $\boldsymbol{Z}_{m,(l_i+l)}$ and $\boldsymbol{Z}_{m',(l_i+l')}$ are independent and identically distributed, i.e., $\mathbb{E} \{\boldsymbol{Z}_{m,(l_i+l)}\boldsymbol{Z}^*_{m',(l_i+l')}\}=\delta(m-m')$.
Therefore, it is sufficient to demonstrate that the elements $\left\{\tilde{\boldsymbol{z}}_n\right\}_m$ are independent and identically distributed (i.i.d.):
\begin{equation}
\small
\label{vzz}
\begin{aligned}
\mathbb{E}\left\{\left\{\tilde{\boldsymbol{z}}_n\right\}_m\left\{\tilde{\boldsymbol{z}}_{n'}\right\}^*_{m'}\right\} 
=& \sum_{l=1}^{L} \sum_{l'=1}^{L} \mathbb{E}\left\{ \boldsymbol{Z}_{m,(l_i+l)} \boldsymbol{Z}^*_{m',(l_i+l')} \right\}\\
&\cdot\boldsymbol{S}^*_{n,l}e^{-j2\pi \mu_i l T_p} \boldsymbol{S}_{n',l'}e^{j2\pi \mu_i l' T_p} \\
=& \sum_{l=1}^{L} \sum_{l'=1}^{L} \delta(m-m')\delta(l-l') \\
=& \delta(m-m')\delta(n-n')
\end{aligned}
\end{equation}
\end{IEEEproof}

Based on Lemma 2, the likelihood function of $\theta_i$ is obtained as
\begin{equation}
\label{pdfR_i}
    p\left (  \boldsymbol{r}_i|\theta _i \right ) 
   =\frac{1}{(L \pi \sigma _z^2)^{N_sN_r}} \exp\{-\frac{1}{2\sigma_z^2}\bar{\boldsymbol{r}}_i^H\bar{\boldsymbol{r}}_i\}
\end{equation}
where $\bar{\boldsymbol{r}}_i=\boldsymbol{r}_i-L \beta_i (\boldsymbol{F}^T \otimes \boldsymbol{b}(\theta_i)) \boldsymbol{a}^*(\theta_i)$.

\subsection{B-Domain Tracking}
\label{BeamDomainTracking}
B-Domain tracking also comprises two components, i.e., B-Domain state prediction and B-Domain state update, where the state prediction is categorized into hard and soft predictions.
In time slot $\eta$, the B-Domain soft state prediction has been defined in \eqref{pPi}.
Then, the hard state prediction is given by 
\begin{equation}
\label{hardprediction}
    \hat{\theta}_i[\eta+1|\eta] =\arg \max \operatorname{Pr}(\theta_i[\eta+1|\eta] ).
\end{equation}
Based on the B-Domain soft state prediction $\boldsymbol{p}({\theta}_i[\eta |\eta-1] )$ in the $[\eta-1]$-th time slot and new measurement $\boldsymbol{R}_i[\eta]$ in (\ref{R_j2}), we can update the AoA of the $i$-th path in $[\eta]$-th time slot employing the maximum a posteriori (MAP) principle, i.e.,
\begin{equation}
\small
\label{MAP}
\hat{\theta}_i[\eta]=\max_{\theta_i[\eta]} p(\theta_i[\eta]|\boldsymbol{R}_i[\eta]) \propto \max_{\theta_i[\eta]} p\left(\boldsymbol{r}_i[\eta]|\theta_i[\eta]\right)p(\theta_i[\eta|\eta-1]),   
\end{equation}
where $p(\theta_i[\eta|\eta-1])= \frac{1}{N_\theta}$ for the non-stationary and unpredicatable transition $(\uppercase\expandafter{\romannumeral3})$ case and \eqref{MAP} degenerates to the maximum likelihood estimation.
Note that AoA estimation has been extensively studied in the field of both signal processing and wireless communication and various well-developed algorithms can be employed to solve \eqref{MAP}, such as the subspace approach \cite{MUSIC} and compressed sensing techniques \cite{yang2021newlowrankoptimization}.
In this work, we focus more on the dual-domain tracking framework and thus simply adopt an exhaustive search method to solve \eqref{MAP} as only one angular parameter is involved in \eqref{MAP}.

\section{CKM-assisted Predictive BF}
\label{predictivebeamform}
In this section, we first derive the CRB of multipath AoA estimation and then design predictive BF for the subsequent time slot to minimize the maximum CRB for AoA estimation of each path, exploiting the channel \textit{a priori} information obtained from the previous time slot.
It is worth noting that accurate B-Domain tracking benefits both target tracking and communication, which motivates us to design predictive BF for optimizing the B-Domain tracking performance in the next time slot.
The trade-off between dual-domain tracking and communication performance is a critical issue of the considered system.
However, C-Domain tracking performance depends on B-Domain tracking performance and thus could be a complicated function of predictive BF strategy.
Moreover, optimizing communication performance needs to consider the uncertainty of predicted CSI based on the predicted AoA for each path, which would lead to a complicated optimization problem.
As a first attempt to propose a dual-domain tracking framework, this paper only considers the B-Domain tracking performance in predictive BF design and the optimization of the trade-off between sensing and communication in the context of dual-domain tracking is left for our future work.

\subsection{CRB for Multi-path AoA Estimation}

CRB serves as a lower bound for the variances of any unbiased estimators \cite{kay1993fundamentals}, which can characterize the multipath AoAs estimation performance.
For each path, given the likelihood function in (\ref{pdfR_i}), the Fisher information (FI) \cite{kay1993fundamentals} of unknown parameter $\theta_i$ is given by $\mathcal{J}\left(\theta_i\right)=- \mathbb{E}\left(\frac{\partial^2 \ln p\left(\boldsymbol{r}_i \mid \theta_i\right)}{\partial \theta_i^2}\right) =
\frac{L^2 |\beta _i|^2}{\sigma_z^2} \boldsymbol{j}(\theta_i)^H \boldsymbol{j}(\theta_i)$,
where $\boldsymbol{j}(\theta_i)= (\boldsymbol{F}^T \otimes \frac{\partial\boldsymbol{b}(\theta _i)}{\partial \theta _i}) \boldsymbol{a}^*(\theta_i)$, and the derivative of the steering vector $\boldsymbol{b}(\theta_i)$ with respect to $\theta_i$ is given by ${\partial\boldsymbol{b}(\theta _i)}/{\partial \theta _i}=j \pi \cos(\theta_i-{\pi}/{2}) \operatorname{diag}( 0,\ldots ,N_r-1 ) \boldsymbol{b}(\theta _i)$.
Based on the definition of CRB \cite{kay1993fundamentals}, the mean squared error (MSE) of AoA estimation is lower bounded by the inverse of $\mathcal{J}\left(\theta_i\right)$, which is given by
\begin{equation}
\label{MSE_CRB}
    \mathbb{E}[(\hat{\theta}_i-\theta_i)^2] \geq {1}/{\mathcal{J}(\theta_i)}  \triangleq \operatorname{CRB}\left(\theta_i\right).
\end{equation}

\textit{Remark 2: }The posterior Cramér-Rao bound (PCRB), which incorporates both the measured data and the \textit{a priori} information of $\theta_i$, is commonly used to characterize the estimation performance in beam tracking frameworks \cite{liuRadarAssistedPredictiveBeamforming2020a}. 
However, as discussed in Fig. \ref{fig_muitpath}, \textit{a priori} information might be unavailable for the third case of the condition of beam transition, i.e., non-stationary and unpredictable transition.
Moreover, even for the first two types of AoA variation, the prior information obtained from TPM is numerical and thus cannot be integrated into PCRB derivation.
Consequently, we adopt CRB as multipath AoA estimation performance metric.
Note that in the $[\eta]$-th time slot, CRB in \eqref{MSE_CRB} depends on ground truth value of $\theta_i$ in the $[\eta+1]$-th time slot, which is unavailable.
However, the B-Domain hard state prediction in \eqref{hardprediction} provides an AoA prediction $\hat{\theta}_i[\eta+1|\eta]$ which can be substituted into \eqref{MSE_CRB} to compute CRB.
We note that this is one way to embed \textit{a prior} information of $\theta_i$ into its predictive estimation error, as commonly adopted in the literature \cite{liuRadarAssistedPredictiveBeamforming2020a}.

\subsection{Predictive BF Design}

In $[\eta]$-th time slot, the predictive BF design to minimize the maximum predictive AoA estimation error $\mathrm{CRB}(\theta_i[\eta+1|\eta])$ is formulated as the following optimization problem
\begin{equation}
\label{problem1}
\small
  \mathop{\mathrm{minimize}}_{ \boldsymbol{F}[\eta+1|\eta]} \max_{i} \ \{\mathrm{CRB}(\theta_i[\eta+1|\eta])\} \ \text{s.t.}||\boldsymbol{F}[\eta+1|\eta]||_F^2  \le P_{\text{t}}.
\end{equation}
where $\boldsymbol{F}[\eta+1|\eta]\in \mathbb{C}^{N_t\times N_s}$ is the predictive BF matrix and $P_{\text{t}}$ is the total transmit power of the RSU.
Utilizing (\ref{MSE_CRB}), the predictive BF design problem can be rewritten as: 
\begin{equation}
\label{problem56}
\small
\mathop{\mathrm{maximize}}_{ \boldsymbol{F}[\eta+1|\eta]} \min_{i} \  \{\mathcal{J}(\theta_i[\eta+1|\eta])\} \ \text{s.t. }   \left \|\boldsymbol{F}[\eta+1|\eta]\right \| _F^2  \le P_{\text{t}}.
\end{equation}
Note that problem (\ref{problem56}) is difficult to be solved globally due to the non-convex objective function. 
As a compromise, we propose a two-step optimization scheme with beam selection and multi-beam power allocation.
In particular, the predictive BF matrix $\boldsymbol{F}[\eta+1|\eta]$ is defined as the product of an analog beam steering matrix $\boldsymbol{A}[\eta+1|\eta] \in \mathbb{C}^{N_t \times N_s}$ and a diagonal power allocation matrix $\boldsymbol{\Gamma}[\eta+1|\eta]\in \mathbb{R}^{N_s \times N_s}$, i.e., $\boldsymbol{F}[\eta+1|\eta]= \boldsymbol{A}[\eta+1|\eta] \boldsymbol{\Gamma}[\eta+1|\eta]$, where $\boldsymbol{\Gamma}[\eta+1|\eta]\triangleq \operatorname{diag} (\sqrt{\gamma_1[\eta+1|\eta]},\cdots,\sqrt{\gamma_{N_s}[\eta+1|\eta]})$, and $\gamma_{n_s}[\eta+1|\eta]$ denotes the power allocated to the $n_s$-th beam in the $[\eta+1]$-th slot. 
\subsubsection{Beam Selection}
The steering matrix $\boldsymbol{A}[\eta+1|\eta]$ is designed based on the B-Domain hard state prediction and its $n_s$-th column is given by
\begin{equation}
\label{A_theta_i}
    \boldsymbol{A}_{:,n_s}[\eta+1|\eta]= \boldsymbol{a}(\hat{\theta}_{n_s}[\eta+1|\eta])\in \mathbb{C}^{N_t\times1} , \forall n_s,
\end{equation}
where $\hat{\theta}_{n_s}[\eta+1|\eta]$ is the predicted AoA for the $n_s$-th path at the $[\eta]$-th time slot in \eqref{hardprediction}.
If $N_s < P$, the predicted AoAs with higher channel path gains should be selected for BF. 
Conversely, if $N_s \ge P$, the predicted AoAs for all $P$ paths can be selected.
This paper proposes to select the strongest paths to improve the dual-domain tracking performance. 
For B-Domain tracking, selecting the strongest paths typically results in higher received SNR, leading to more reliable AoA estimates and prediction, which are crucial for maintaining beam alignment.
For C-Domain tracking, the Fisher information for target state estimation is dominated by the strongest paths. 
Specifically, the LoS path enables an analytical measurement model derived from its deterministic geometric relationship associated with the target state, providing high-quality measurements, while the strongest NLoS paths become critical in the LoS-absent case, enabling continuous tracking with the assistance of the CKM, which serves as a numerical measurement model.

\subsubsection{Power Allocation}
The second step is to acquire the optimal power allocation matrix $\boldsymbol{\Gamma}[\eta+1|\eta]$ based on the given steering matrix $\boldsymbol{A}[\eta+1|\eta]$. 
The objective function of \eqref{problem56} given $\boldsymbol{A}[\eta+1|\eta]$ is given by \eqref{eta_J_theta_i} at the top of next page, where $a_i \triangleq \beta_i \sin \left(\theta_i\right),\forall i $ for simplicity.
\begin{figure*}[!t]
\normalsize
\small
\begin{equation}
\label{eta_J_theta_i}
\begin{aligned}
     \mathcal{J}(\theta_i [\eta+1|\eta]) 
     & =\frac{L^2 \pi^2 |a_i[\eta+1|\eta]|^2 }{\sigma_z^2} 
        \left\{ \left[ \left( \boldsymbol{\Gamma}^T[\eta+1|\eta] \boldsymbol{A}^T[\eta+1|\eta] \right) \otimes \left( \operatorname{diag}(0,\ldots,N_r-1)\boldsymbol{b}(\theta_i[\eta+1|\eta]) \right) \right] \boldsymbol{a}^*(\theta_i[\eta+1|\eta]) \right\}^H \\
        & \quad \cdot \left\{ \left[ \left( \boldsymbol{\Gamma}^T[\eta+1|\eta] \boldsymbol{A}^T[\eta+1|\eta] \right) \otimes \left( \operatorname{diag}(0,\ldots,N_r-1)\boldsymbol{b}(\theta_i[\eta+1|\eta]) \right) \right] \boldsymbol{a}^*(\theta_i[\eta+1|\eta]) \right\}.
\end{aligned}
\end{equation}
\hrulefill
\end{figure*}
Since we focus on the power allocation for each time slot, the time index in (\ref{eta_J_theta_i}) is omitted, i.e., $\beta _i[\eta+1|\eta]$, $ \boldsymbol{\Gamma}[\eta+1|\eta]$, $ \boldsymbol{A}[\eta+1|\eta]$, $\theta_i[\eta+1|\eta]$ and $a_i[\eta+1|\eta]$ are represented by $\beta _i$, $ \boldsymbol{\Gamma}$, $ \boldsymbol{A}$, $\theta_i $ and $a _i$, respectively. 
Using the mixed-product property of the Kronecker product, we can obtain
$\mathcal{J}(\theta_i) = ({L^2  \pi^2 a_i^2}/{\sigma_z^2}) \boldsymbol{c}^H(\theta_i) ( \boldsymbol{\Gamma}^T \otimes \operatorname{diag}(0,\ldots,N_r-1) )^H  ( \boldsymbol{\Gamma}^T \otimes \operatorname{diag}(0,\ldots,N_r-1) ) \boldsymbol{c}(\theta_i)$, where $\boldsymbol{c}(\theta_i) \triangleq  ( \boldsymbol{A}^T \otimes \boldsymbol{b}(\theta _i) )  \boldsymbol{a}^*(\theta_i) \in \mathbb{C}^{N_s N_r\times 1}$ can be calculated based on (\ref{A_theta_i}). 
Applying the mixed-product property of the Kronecker product again, yields $\mathcal{J}(\theta_i) = ({L^2  \pi^2 a_i^2}/{\sigma_z^2}) \boldsymbol{c}^H(\theta_i)[\boldsymbol{\Gamma} \boldsymbol{\Gamma}^H \otimes \operatorname{diag}^2(0,\ldots, N_r-1) 
] \boldsymbol{c}(\theta_i)$.
Ultimately, $\mathcal{J}\left(\theta_i\right)$ in (\ref{eta_J_theta_i}) can be rewritten as $\mathcal{J}\left(\theta_i\right)=({L^2 \pi^2 a_i^2 }/{\sigma_z^2}) \sum_{{n_s}=1}^{N_s}\gamma_{n_s}\sum_{g=1}^{N_r}|\{\boldsymbol{c}(\theta_i)\}_{({n_s}-1) N_r+g}|^2(g-1)^2$, where $\{\boldsymbol{c}(\theta_i)\}_{k}$ denotes the $k$-th element of the vector $\boldsymbol{c}(\theta_i)$. 
Further introducing auxiliary variables $t \ge 0$ to bound the minimum value of $\mathcal{J}(\theta_i)$, problem (\ref{problem56}) can be equivalently rewritten as
\begin{equation}
\small
\label{finalOP}
\begin{aligned}
\min_{t, \gamma_1, \ldots, \gamma_{N_s}} & t \\
\text{s.t. } & t \leq a_i^2 \sum_{{n_s}=1}^{N_s} \gamma_{n_s} c_{i,{n_s}}, \sum_{{n_s}=1}^{N_s} \gamma_{n_s} \leq \frac{P_{\text{t}}}{N_t} ,\gamma_{n_s} \geq 0, \forall i,\forall{n_s},
\end{aligned}
\end{equation}
where $c_{i,{n_s}}= \sum_{g=1}^{N_r} |\{\boldsymbol{c}(\theta_i)\}_{({n_s}-1)N_r+g} |^2(g-1)^2$.
After the proposed series of transformations, it can be verified that the power allocation problem \eqref{finalOP} is convex and can be solved using off-the-shelf numerical convex programming tools, such as CVX \cite{cvx}. 
Problem \eqref{finalOP} is a linear programming problem with $N_s + 1$ variables and $(P \times N_s) + 1$ linear constraints, which can be solved in polynomial time using interior-point methods with a complexity of $O((P + N_s)^{3.5})$ \cite{Karmarkar1984LP}.
For typical parameter settings in our simulations, this process requires fewer than $10^4$ floating-point operations, which can typically be executed within a few milliseconds, depending on the processor used, and is well below the slot duration $\Delta T$.

\textit{Remark 3: }It is worth noting that the beam selection step defines both the optimization space and landscape for the subsequent power allocation design and thus directly determines the power distribution among the selected beams.
In \eqref{finalOP}, the beam selection determines the number of optimization variables $\gamma_1, \ldots, \gamma_{N_s}$ and hence the problem dimensionality.
Power allocation then optimizes the power distribution among the selected beams to minimize the maximum AoA estimation error across paths.
Furthermore, the coefficients $c_{i,{n_s}}$ of \eqref{finalOP} depend jointly on the transmit and receive steering vectors of the RSU as well as the beam selection matrix $\boldsymbol{A}$.
In turn, power allocation affects the updated and predicted AoA in the next time slot and then indirectly affects the beam selection.

\textit{Remark 4: }The formulation of \eqref{problem1} can be generalized to a joint optimization problem by introducing a communication constraint, such as constraints on the minimum achievable rate or maximum outage probability, based on the predicted CSI which implies the sensing-communication tradeoff specifically.
A more stringent communication constraint (e.g., a higher minimum rate requirement) tends to allocate more transmit power to communication, which reduces the power concentrated along the predicted AoAs and consequently increases the CRB for AoA estimation. 
Conversely, a less stringent communication constraint allows more power concentrated on the predicted AoAs, thereby minimizing AoA estimation errors.
A joint optimization of predictive beamforming for both sensing and communication represents an important direction for our future work.

\section{Simulation Results}

In this section, we present the numerical results to evaluate the performance of the proposed CKM-assisted dual-domain tracking and predictive BF framework. 

\subsection{Simulation Parameters And Baseline Schemes}

The RSU is positioned at the origin and the vehicle starts from the coordinates $[-20 \mathrm{~m}, 10 \mathrm{~m}]^T$ and moves along the road with an initial speed of $10 \mathrm{~m/s}$. 
The time of interest is $T_{\max}=4 \mathrm{~s}$ with a slot duration of $\Delta T = 0.02 \mathrm{~s}$. 
The carrier frequency $f_c$ is set to $30 \mathrm{~GHz}$ and the sample interval $T_p$ is $10^{-8}$ seconds. 
Both the maximum delay and Doppler shift indices are set as $100$. 
The total power transmit $P_{\text{t}}$ is $16 \mathrm{~W}$ and the angular grid size is set as $N_\theta=7200$. 
For the C-Domain state evolution noises, we set the standard deviations for position $\sigma_{q_x}, \sigma_{q_y}$ are $10^{-3}\mathrm{~m}$, while $\sigma_v$ is $10^{-3}\mathrm{~m/s}$. 
The standard deviations for the delay and Doppler shift measurement noise are $\sigma_\tau= 10^{-8}\mathrm{~s}$ and $\sigma_\mu=20\mathrm{~Hz}$, respectively. 
The standard deviation of angle measurement noise is $\sigma_{\cos(\theta)}=0.01$ and AoA variation model parameters are $\xi=0.8$, $\epsilon =\hat{v}\sin(\hat{\theta}_1)N_\theta/20$, and $\sigma_{\text{CKM}}=10^{-3}$.
Each cumulative distribution function (CDF) curve is obtained from 100 Monte Carlo runs.

The ray tracing toolbox in MATLAB is leveraged to generate the ground-truth channel information \cite{matlab_raytracing}. 
A CKM is designed to record the $P=2$ strongest channel paths, denoted as \textbf{path 1} and \textbf{path 2}, respectively\footnote{Note that we consider two paths since the simulated environment exhibits two dominant propagation paths that capture the major multipath structure of the considered scenario.
Therefore, we set $P=2$ in our simulations.
Moreover, the proposed scheme is applicable to the case with more propagation paths.}. 
The CKM is constructed adopting the IDW of the $K$ nearest neighbors (KNN) approach with a limited set of 3200 ray tracing data samples, as commonly adopted in the literature  \cite{ChilesDelfiner2009Geo,CoverHart1967NN,liChannelKnowledgeMap2022}. 
The selected sample size achieves a spatial resolution of approximately $0.1$ m in the considered area, which is comparable to the vehicle displacement between consecutive slots ($v \Delta T=0.2$ m).
The temporally correlated blockage caused by static and large size obstacles in the propagation environment is assumed to emerge between the $140$-th and $175$-th time slots, whereas random blockage events caused by transient obstacles appear with a certain blockage probability.
Note that NLoS path is more likely to be blocked than LoS path in practice \cite{blocked}.
Therefore, we can define the blockage probability of the LoS path as $p_{\text{blk}}$ and the block probability of an NLoS path as $1-(1-p_{\text{blk}})^2$ since an NLoS path can be treated as the concatenation of two LoS paths.
We note that LoS-absent happens when the LoS path is blocked. 
Besides, the blockage of NLoS path is not treated as a LoS-absent case.

\begin{figure}[!t]
\centering
\includegraphics[width=2.5 in]{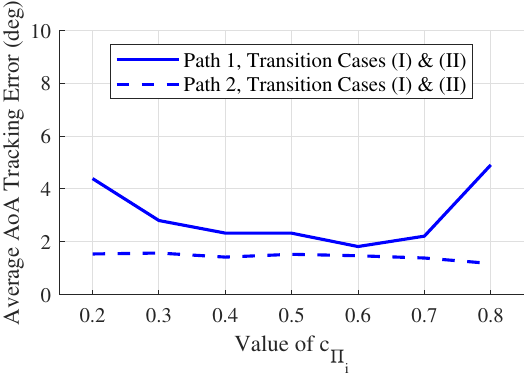}
\caption{Average AoA tracking errors with respect to $c_{\Pi_i}[\eta]$ for all paths and slots with $\sigma_z^2=10^{-6} \mathrm{~W}$, $N_t=N_r=32$, and $p_{\text{blk}}=0.15$.}
\label{randomError_cPi}
\end{figure}
For the stationary transition case, the TPM constant is set to $c_{\Pi_i}[\eta] = 0.6$ for all paths and time slots. 
This value was determined based on the observed relationship between $c_{\Pi_i}$ and tracking accuracy observed in Fig. \ref{randomError_cPi}, which shows that the average AoA tracking error first decreases and then increases as $c_{\Pi_i}$ grows, particularly for \textbf{path 1}. 
The initial improvement in AoA tracking accuracy arises because a moderate $c_{\Pi_i}$ leverages the CKM prior in the angular TPM in \eqref{Pi} to enhance multipath AoA estimation.
However, the CKM prior is inherently imperfect, since its input position is predicted rather than measured, which introduces a mismatch between the predicted and actual AoA. 
The C-Domain tracking error, together with the uncertainty of CKM construction, further reduces B-Domain tracking accuracy.
Therefore, the average AoA tracking error increases with further increasing $c_{\Pi_i}$.

In this section, a baseline method described in \cite{zhaoSensingAssistedPredictiveBeamforming} is considered to demonstrate the benefits of the proposed framework. 
In particular, the baseline algorithm in \cite{zhaoSensingAssistedPredictiveBeamforming} integrates the result of LoS path blockage detection with an EKF-based tracking framework where the position, speed, and LoS path AoA are jointly tracked.
Note that the baseline is designed solely based on the LoS channel model and fails to account for the information of NLoS paths.
Therefore, we assume that the baseline scheme can perfectly separate the received signal of LoS path in the LoS-present case, which is utilized for tracking.
If the LoS path is absent, the sensing receiver ignores the observation of the NLoS path and performs state prediction only to predict the position, speed, and LoS path AoA of the vehicle as considered in \cite{zhaoSensingAssistedPredictiveBeamforming}.

\subsection{Tracking Performance}
\begin{figure}[!t]
\centering
\includegraphics[width=3.1 in]{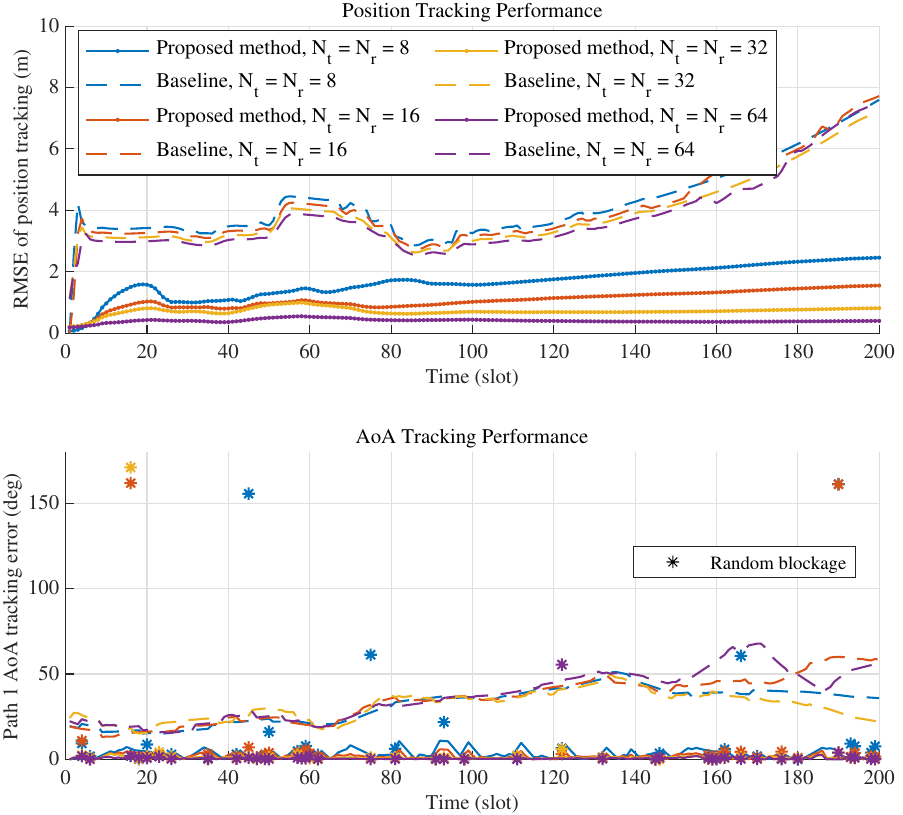}
\caption{Tracking performance for the proposed dual-domain framework versus the time index with $\sigma_z^2=10^{-9} \mathrm{~W}$ and $p_{\text{blk}}=0.15$.}
\label{trackinghighSNR}
\end{figure}
\begin{figure}[!t]
\centering
\includegraphics[width=3.1 in]{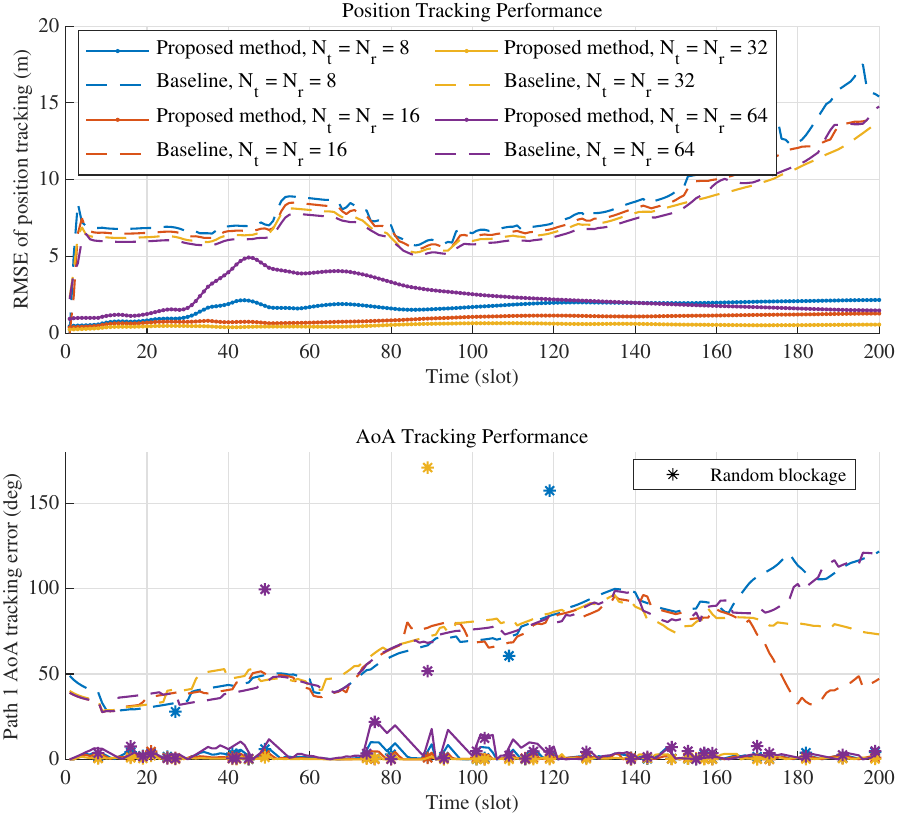}
\caption{Tracking performance for the proposed dual-domain framework versus the time index with $\sigma_z^2=10^{-6} \mathrm{~W}$ and $p_{\text{blk}}=0.15$.}
\label{trackinglowSNR}
\end{figure}
\begin{figure}[!t]
\centering
\subfloat[$\sigma_z^2=10^{-9} \mathrm{~W}$]{\includegraphics[width=1.5 in]{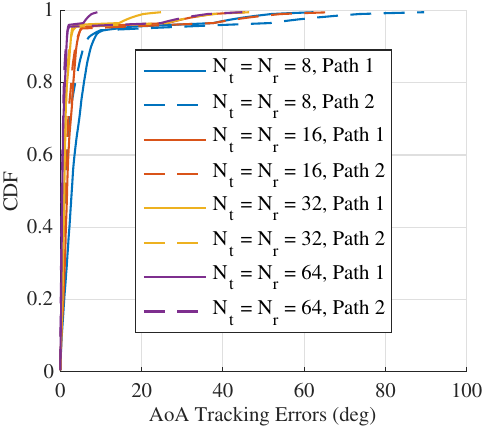} \label{CDFhighSNR}}
\hspace{0cm}
\subfloat[$\sigma_z^2=10^{-6} \mathrm{~W}$]{\includegraphics[width=1.5 in]{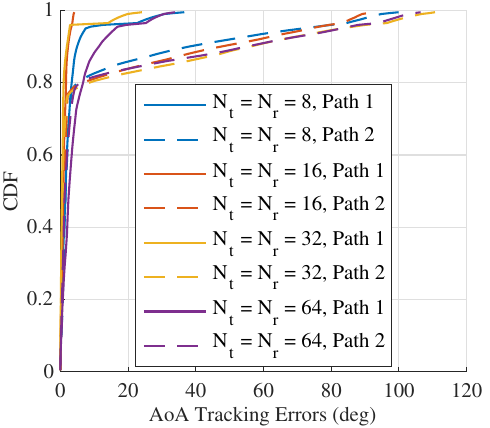} \label{CDFlowSNR}}
\caption{CDF of B-Domain multipath AoA tracking errors with $p_{\text{blk}}=0.15$ at different $\sigma_z^2$ values.}
\end{figure}
Fig. \ref{trackinghighSNR} illustrates the dual-domain tracking performance in terms of root mean squared error (RMSE) for the vehicle position and the absolute error for AoA estimation for \textbf{path 1}.
In contrast to the baseline scheme, the proposed dual-domain EKF tracking framework ensures continuous target tracking and achieves a higher tracking accuracy even during the prolonged LoS-absent period.
This is because our proposed scheme comprehensively models of the beam transition conditions and effectively utilizes the CKM to extract NLoS path information for positioning.
Moreover, it can be observed that all the position tracking RMSE curves first increase and then decrease/stabilize with the evolution of time. 
This behavior is due to that the tracking accuracy of EKF algorithm requires several iterations to stabilize.
Additionally, when \textbf{path 1} is randomly blocked, the AoA of the strongest NLoS path changes unpredictably leading to beam misalignment and spikes in AoA tracking error indicated by asterisks in Fig. \ref{trackinghighSNR}.
The AoA tracking errors are mostly concentrated near 0°, indicating that the B-Domain tracking maintains a high AoA estimation accuracy even under random blockages.
In contrast, a large tracking error happens when the LoS path disappears due to random blockage and the induced path lies far from the predicted beam direction, even occasionally approaching 180°. 
This is because a sudden change in the B-Domain state leads to severe beam misalignment and a significantly reduced received SNR for the induced path.
Fortunately, these spikes do not significantly affect overall position tracking accuracy, as the measurement models in the C-Domain for LoS-present and LoS-absent cases are different.
Note that the multipath AoA estimation accuracy has not been provided in Fig. \ref{trackinghighSNR} since the baseline only tracks the AoA of the LoS path.
In the first few time slots, the AoA of LoS path in the following time slots of the baseline scheme is larger than that of the proposed framework because the MAP-based update principle proposed in Section \ref{BeamDomainTracking} effectively incorporates \textit{a priori} information when echo signal measurements are unreliable or have low SNR.
In contrast, large errors in both target and beam tracking for the baseline scheme arise with time slot, even reaching more than $10$ meters of RMSE of position tracking and $50$ degrees of \textbf{path 1} AoA tracking error in the end.
This is because for LoS-absent cases, the echo signal measurement is treated as meaningless in the baseline scheme in \cite{zhaoSensingAssistedPredictiveBeamforming} and relies on the state evolution model to predict the position of the vehicle and AoA of LoS path in the following time slots.
Since the state evolution model adopted may suffer from model mismatch, with the increase of the slot, both the target and beam tracking error scale up.

Fig. \ref{trackinglowSNR} considers a higher noise power to evaluate the performance of the proposed scheme.
When the noise power is higher, a larger antenna array can improve tracking accuracy due to a higher BF gain, but they also increase the likelihood of beam misalignment, as shown in Fig. \ref{trackinglowSNR}. 
For a larger antenna array with $ N_t = N_r = 64$, both target and beam tracking performance become worse when the vehicle is close to the RSU, specifically between $80-100$ slots, even for the cases where \textbf{path 1} has not been blocked.
In fact, the AoA of \textbf{path 1} varies rapidly in these time slots and B-Domain state prediction becomes inaccurate, and thus predictive BF with a narrow beam for a larger antenna array is more susceptible to missing the moving target.

Note that in both Fig. \ref{trackinghighSNR} and Fig. \ref{trackinglowSNR}, baseline methods fail to maintain accurate tracking in the considered complex and high-mobility wireless network due to random path blockage.
The proposed CKM-assisted dual-domain tracking framework outperforms the baseline for two key reasons. 
First, based on the C-Domain tracking results, the CKM provides additional \textit{a priori} information for B-Domain tracking, which improves AoA tracking accuracy, even under poor channel conditions. 
Second, conventional EKF-based target tracking schemes, such as those in prior works \cite{cui2024SeeingNotAlwaysBelievingISACAssistedPredictiveBeamTrackingMultipathChannels,liISACEnabledV2INetworks2023,zhaoSensingAssistedPredictiveBeamforming,yuStatisticalNLOSIdentification2009}, are highly sensitive to model mismatch errors due to discrepancy between the assumed measurement model and the actual propagation environment. 
In contrast, the proposed framework utilizes the CKM to serve as the measurement function in the LoS-absent case in addition to the traditional geometric-based measurement model in the LoS-present case.

\begin{figure}[!t]
\centering
\includegraphics[width=3 in]{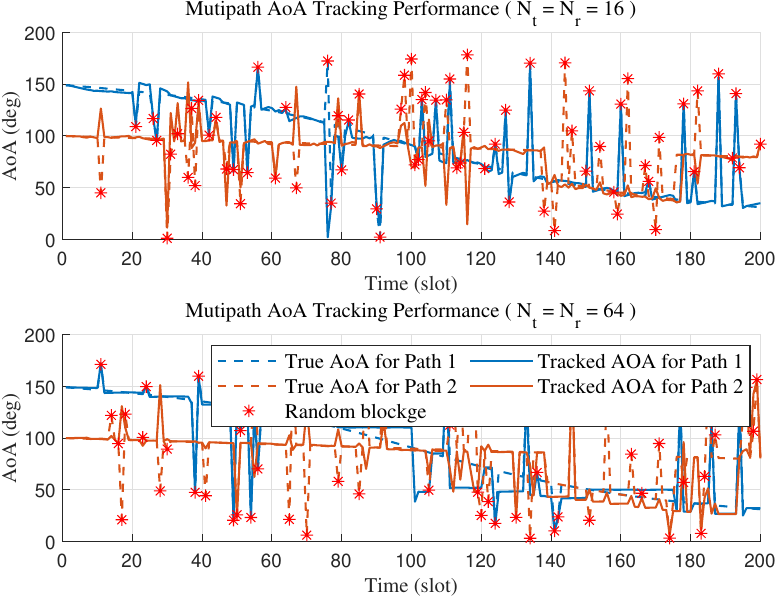}
\caption{Multipath AoA tracking performance with $\sigma_z^2=10^{-6} \mathrm{~W}$.}
\label{multitracking}
\end{figure}
Furthermore, Fig. \ref{CDFhighSNR} and Fig. \ref{CDFlowSNR} illustrate the CDF of AoA tracking error for \textbf{path 1} and \textbf{path 2} for different configurations of $N_t$ and $N_r$, corresponding to Figs. \ref{trackinghighSNR} and \ref{trackinglowSNR}, respectively.
Due to the lower noise power $\sigma_z^2$, Fig. \ref{CDFhighSNR} consistently exhibits better AoA estimation accuracy than Fig. \ref{CDFlowSNR} for all paths and antenna configurations.
In contrast, Fig. \ref{CDFlowSNR} reveals that a higher measurement noise power significantly impacts the AoA tracking accuracy, particularly for \textbf{path 2}. 
This degradation is most pronounced during non-stationary and unpredictable beam state transitions, where the AoA estimation relies solely on the current observed signal and thus is more sensitive to noise-induced distortions. 
Consistent with the observations in Fig. \ref{trackinglowSNR}, increasing the number of antennas is not always beneficial for AoA tracking, particularly for \textbf{path 2} (NLoS path) and the case with a higher noise power.
As the NLoS path has a higher random blockage probability, a smaller antenna array with a wider predictive beam has a higher probability of covering the bursty changed AoA of \textbf{path 2} and thus may provide a high-quality signal measurement in the B-Domain.
It is important to note that $p_{\text{blk}} = 0.15$ indicates that $85\%$ of the AoA estimates for \textbf{path 1} are obtained with the assistance of \textit{a priori} information and $72.25\%$ of the AoA estimates for \textbf{path 2} are supported with \textit{a priori} information.
In both Figs. \ref{CDFhighSNR} and \ref{CDFlowSNR}, the AoA tracking error is lower than $10$ degrees with the assistance of \textit{a priori} information.
This indicates that incorporating \textit{a priori} information significantly improves AoA tracking accuracy, even under low SNR conditions, especially for the NLoS path.
This finding underscores the robustness of our proposed approach in maintaining accurate angle estimates, even in the presence of significant noise and high-speed moving target.
Fig. \ref{multitracking} further corroborates the analysis above, demonstrating that while larger antenna arrays can provide a better AoA tracking accuracy in stationary or predictable cases, they are more sensitive to unexpected beam state transition in the propagation environment.

\begin{figure}[!t]
\centering
\includegraphics[width=2.5 in]{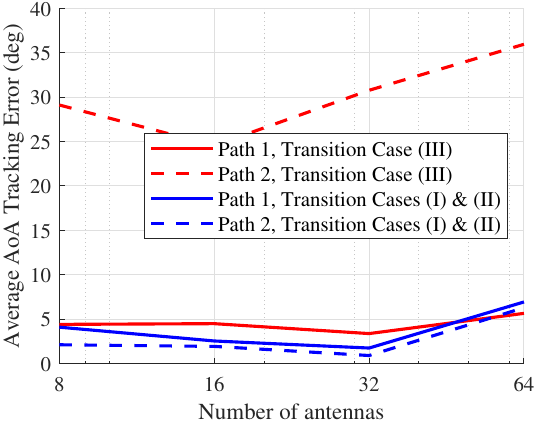}
\caption{Average AoA tracking errors with respect to the number of antennas.}
\label{randomError}
\end{figure}
Fig. \ref{randomError} further demonstrates that non-stationary and unpredictable transition $(\uppercase\expandafter{\romannumeral3})$ degrades the AoA tracking performance compared to the other two types of beam state transition, particularly for NLoS path and the case with a larger antenna array.
In fact, in the third type of beam state transition, without \textit{a prior} information of channel and relying solely on the received signal, it is challenging to estimate the AoA for the NLoS path due to its much higher propagation loss than that of LoS path.
Moreover, in this case, the RSU struggles to predict the beam and predictive BF may suffer from the beam misalignment with a higher probability for a larger antenna array.

\subsection{Predictive BF Performance}
\begin{figure}[!t]
\centering
\includegraphics[width=2.5 in]{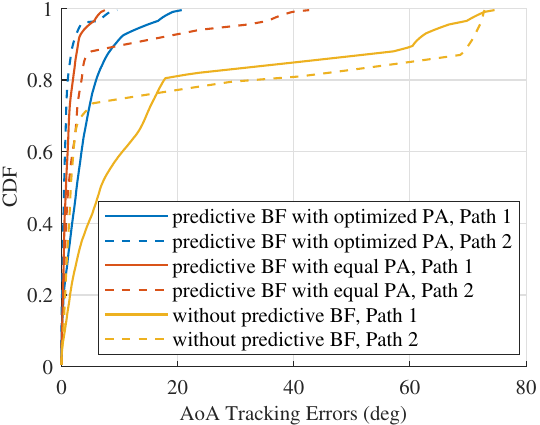}
\caption{AoA tracking errors of different predictive BF schemes.}
\label{predictiveBF}
\end{figure}
Fig. \ref{predictiveBF} illustrates that incorporating predictive BF and optimizing power allocation (PA) significantly improves multipath AoA tracking accuracy. 
For \textbf{path 1}, Fig. \ref{predictiveBF} shows that BF with equal PA achieves the highest tracking accuracy, followed by predictive BF with optimized PA, and thirdly, the case without predictive BF. 
For \textbf{path 2}, predictive BF with optimized PA achieves the highest AoA tracking accuracy, followed by predictive BF with equal PA, and lastly, the case without predictive BF.
The optimization of power allocation balances the sensing performance across different propagation paths.
By adopting the power distribution among multiple beams, the proposed method effectively improves the AoA tracking performance of the path with a lower path gain. 
This is particularly important for NLoS paths, where predictive BF with optimized PA allows for enhanced tracking performance by compensating its propagation loss.
Overall, the combination of predictive BF and optimized PA offers a more robust solution for multipath AoA tracking in complex environments.

\section{Conclusion}

In this paper, we proposed a CKM-assisted dual-domain tracking and predictive BF framework for high-mobility wireless networks. 
In the C-Domain, we proposed an EKF-based target tracking method that is effective for both LoS-present and LoS-absent conditions. 
In the B-Domain, we developed an angular TPM that combines the channel temporal correlation and \textit{a priori} information from CKM for each path to assist AoA tracking.
Moreover, we analyzed the CRB of AoA estimation and proposed a predictive BF and power allocation design to minimize the maximum AoA estimation error across multiple paths for the next time slot.
Simulation results confirm that CKM significantly enhances both the target and beam tracking accuracy, especially in complex high-mobility wireless communication networks. 
Additionally, incorporating predictive BF and optimized power allocation further improves AoA tracking performance, especially for NLoS paths with high propagation loss.
This framework can be expanded to multi-user and multi-UAV networks, accommodating the simultaneous movement of multiple nodes.
The proposed technique is a pivotal advancement for the development of intelligent transportation systems, offering improved communication reliability and efficiency in 6G networks.

\appendices

\bibliographystyle{IEEEtran}
\bibliography{main}
\end{document}